\documentclass[useAMS,usenatbib]{mn2e}
\usepackage[english]{babel}

\usepackage{fancyhdr}
\usepackage{amsfonts}
\usepackage{amsmath}
\usepackage{amssymb}
\usepackage{multicol}
\usepackage{layout}
\usepackage{graphicx}
\usepackage{multirow}

\usepackage{times}
\usepackage{natbib}

\newif\ifAMStwofonts
\AMStwofontstrue

\graphicspath{{./fig/}}

\title[CoDECS raytracing simulations]{Raytracing simulations of coupled dark energy models}

\author[Pace et~al.]{Francesco Pace$^{1}$\thanks{francesco.pace@port.ac.uk}, Marco Baldi$^{2,3,4}$, Lauro
Moscardini$^{2,3,4}$, David Bacon$^1$, Robert Crittenden$^1$\\
$^1$ Institute of Cosmology and Gravitation, University of Portsmouth, Dennis Sciama Building, Portsmouth, PO1 3FX,
U.K.\\
$^2$ Dipartimento di Fisica e Astronomia, Universit\`a di Bologna,
Viale Berti Pichat 6/2, I-40127 Bologna, Italy\\
$^3$ INFN, Sezione di Bologna, Viale Berti Pichat 6/2, I-40127 Bologna, Italy\\
$^4$ INAF, Osservatorio Astronomico di Bologna, Via Ranzani 1, I-40127 Bologna, Italy}

\date{Accepted 2014 November 25, Received 2014 October 26; in original form 2014 August 2}

\pagerange{\pageref{firstpage}--\pageref{lastpage}} \pubyear{2013}

\begin{document}
\label{firstpage}
\maketitle

\begin{abstract}
Dark matter and dark energy are usually assumed to couple only gravitationally. 
An extension to this picture is to model dark energy as a scalar field coupled directly to cold dark matter. This 
coupling leads to new physical effects, such as a {\it fifth}-force and a time-dependent dark matter particle mass. 
In this work we examine the impact that coupling has on weak lensing statistics by constructing realistic simulated 
weak-lensing maps using raytracing techniques through N-body cosmological simulations. We construct maps for 
different lensing quantities, covering a range of scales from a few arcminutes to several degrees. 
The concordance $\Lambda$CDM model is compared to different coupled dark energy models, described either by an 
exponential scalar field potential (standard coupled dark energy scenario) or by a SUGRA potential (bouncing model). 
We analyse several statistical quantities and our results, with sources at low redshifts are largely consistent with 
previous work on CMB lensing by \cite{Carbone2013}. 
The most significant differences from the $\Lambda$CDM model are due to the enhanced growth of the perturbations 
and to the effective friction term in non-linear dynamics. 
For the most extreme models, we see differences in the power spectra up to 40\% compared to the $\Lambda$CDM 
model. 
The different time evolution of the linear matter overdensity can account for most of the differences, but when 
controlling for this using a $\Lambda$CDM model having the same normalization, the overall signal is smaller due to 
the effect of the friction term appearing in the equation of motion for dark matter particles.
\end{abstract}


\section{Introduction}
A wealth of cosmological probes have confirmed the accelerated expansion of the Universe first inferred with 
observations of Type Ia SNae 
\citep{Riess1998,Schmidt1998,Perlmutter1999,Riess2004,Astier2006,Kowalski2008,Kessler2009,Conley2011,Sullivan2011}.
These include the angular power spectrum of the cosmic microwave background (CMB) fluctuations and the integrated 
Sachs-Wolfe effect
\citep{Jaffe2001,Giannantonio2008,Ho2008,Jarosik2011,Komatsu2011,Sherwin2011,Hinshaw2013,Planck2013_XV,Planck2013_XVI,
Planck2013_XIX}, the number counts of massive galaxy clusters 
\citep{Haiman2001,Allen2004,Allen2008,Wang2004,Vikhlinin2009,Mantz2010,Rapetti2010,Rozo2010,Benson2013}, 
weak lensing 
\citep{Hoekstra2006,Jarvis2006,Fu2008,Schrabback2010,Kilbinger2013}, 
galaxy clustering 
\citep{Percival2001,Tegmark2004a,Cole2005,Guzzo2008,Reid2010,Blake2012,delaTorre2013a}
and baryon acoustic oscillations (BAO) 
\citep{Eisenstein2005,Percival2010,Parkinson2012,Sanchez2014,Veropalumbo2014}.
To explain this acceleration, a new dark component with equation of state $w<-1/3$ has been introduced dubbed dark 
energy (DE). As with dark matter (DM), the DE also does not interact with the electromagnetic field 
\citep[and references therein]{Bertone2005,Bartelmann2010,Astier2012} and its nature is still completely unknown after 
more than a decade of theoretical and observational investigations.

Continuous improvements in observations have led to the definition of a standard model in cosmology; in the 
Concordance Cosmological Model (CCM) the Universe is filled with baryons ($\simeq 5\%$ of the total energy budget), 
dark matter ($\simeq 27\%$) and dark energy ($\simeq 68\%$) \citep{Planck2013_XVI}.  
In its simplest form, the dark energy is assumed to be a cosmological constant,  
characterised by an equation of state ($w=-1$) and energy density constant throughout the whole cosmic history.

Despite its simplicity, the CCM fits virtually all the available observations 
\citep[][and references therein]{Planck2013_XVI}. Nevertheless, the cosmological constant suffers severe problems 
from a theoretical point of view. In particular the actual value of the cosmological constant requires an extreme 
fine-tuning, giving rise to the coincidence \citep{Zlatev1999} and fine tuning problems 
\citep{Weinberg1989,Sahni2000}. This provides motivation to find viable alternatives to overcome these fundamental 
problems, for example considering dynamical dark energy models or modifications to gravity.

Moreover, despite the fact that the cosmological constant scenario can explain most observations at cosmological 
scales, many phenomena at small and intermediate scales indicate possible problems with this simple model. 
These include the lack of luminous satellites in cold dark matter haloes \citep{Navarro1996,BoylanKolchin2011}, the 
observed low baryon fraction in galaxy clusters \citep{Ettori2003,McCarthy2007}, and the high velocities detected in 
the large-scale bulk motion of galaxies \citep{Watkins2009}. 
A simple explanation for these features may be related to the fact that our understanding of the baryonic physics 
at these scales is still very incomplete, but nevertheless it is worth investigating whether alternative models could 
accommodate or diminish the tension between these observations and theory.

One interesting direction is to study interactions in the dark sector between the dark energy and the dark matter 
component. Coupled dark energy models were first introduced by \cite{Wetterich1995} and \cite{Amendola2000} in order 
to alleviate the fine-tuning problem; these have since been studied in some detail 
\citep{Amendola2004,Amendola2007a,Pettorino2008,Amendola2008,diPorto2008,CalderaCabral2009a,CalderaCabral2009b,
Boehmer2010, Koyama2009,LopezHonorez2010,Majerotto2010,Valiviita2010,Baldi2011c,Baldi2011d,Baldi2012,Clemson2012}.
Observational constraints on the interaction strength were obtained using the CMB 
\citep{Bean2008,LaVacca2009,Xia2009}. 
These models have also been investigated using numerical simulations
\citep{Maccio2004,Baldi2010,Li2011a,Li2011c,Li2011b,Baldi2011} which showed 
that significant deviations from the $\Lambda$CDM model have to be expected in the non-linear regime.

In this class of models the role of dark energy is played by a dynamical scalar field and there is a coupling 
describing an exchange of energy-momentum between dark matter and dark energy. While observations put strong 
constraints on the amount of interaction between the baryons and the dark sector \citep{Hagiwara2002}, this is not 
the case for interactions in the dark sector. A consequence of the coupling is the rise of a fifth force that modifies 
the equations of motion of dark matter and significantly affects the evolution of the collapsing structures. To 
account for this in the non-linear dynamics it is necessary to use expensive numerical simulations. 
In this work we make use of the largest available suite of such N-body simulations called {\small CoDECS} 
\citep[COupled Dark Energy Cosmological Simulation;][]{Baldi2012}. These simulations have been used to study the halo 
mass function \citep{Cui2012}, the BAO \citep{Cervantes2012}, the galaxy rotation curves \citep{Baldi2012a}, 
the redshift-space distortions \citep{Marulli2012}, the pairwise infall velocity of colliding clusters 
\citep[][]{Lee2012}, and the gravitational lensing effect \citep{Beynon2012,Carbone2013}.

\cite{Beynon2012} made predictions for the shear correlation function in the non-linear weak lensing regime based on 
CoDECS simulations of three `standard' coupled dark energy models with an exponential potential and a reference 
$\Lambda$CDM model. This work utilised the analytical relation between the matter power spectrum and the shear 
correlation function: the shear power spectrum can be written as an integral along the line of sight of the 
(non-linear) matter power spectrum \citep{Bartelmann2001}. Taking the input matter power spectrum obtained directly 
from the particle distribution in the box, they derived predictions for the shear correlation; they also made 
forecasts for the {\it Dark Energy Survey} (DES)\footnote{http://www.darkenergysurvey.org} and the Euclid 
mission\footnote{http://www.euclid-ec.org} \citep{Laureijs2011,Amendola2013} and showed that it will be possible
to use lensing to distinguish between $\Lambda$CDM and coupled dark energy models at a $4-\sigma$ level.

\cite{Carbone2013} instead performed a raytracing analysis, focusing on CMB lensing rather than lower redshift 
sources. Using the snapshots of the simulated box, the authors constructed deflection angle maps and studied the 
statistical properties of the deflection angle and lensing potential power spectrum. 
They analysed three different models: a reference $\Lambda$CDM model and two different coupled dark energy models; a 
standard scenario with an exponential potential, and a bouncing model described by a SUGRA potential 
\citep{Brax1999}. The authors showed that for the standard scenario, differences with the $\Lambda$CDM model arise 
from the interplay between an enhanced growth and a modified non-linear structure formation, while for the bouncing 
model these two effects make the power of the lensing signal $\approx 10\%$ smaller than for the reference 
$\Lambda$CDM model.

In this work, we extend both previous works \citep{Beynon2012,Carbone2013} with a full numerical analysis of the 
statistical properties of several lensing quantities. 
In particular we analyse the superset of models studied in these two works from a completely numerical point of view, 
basing our work on raytracing simulations. 
An important goal of this work is to validate the semi-analytic method of \cite{Beynon2012} with a full numerical 
approach, and to check whether previous results are in agreement with a full non-linear treatment. 
Due to the different linear evolution in the coupled models, the perturbations have a different normalization of the 
matter power spectrum. In addition, the non-linear dynamics is different from the $\Lambda$CDM model, and effects 
induced by it might not be captured with the semi-analytical treatment. In order to separate the linear 
normalisation differences from the differences in the non-linear physics, we also make a comparison with analytical 
models in the $\Lambda$CDM cosmology.

The paper is organised as follows. In Section~\ref{sect:DEmodels} we briefly describe the main properties of the 
coupled dark energy cosmologies. The corresponding N-body simulations are described in Section~\ref{sect:Nbody}. 
In Section~\ref{sect:lensing} we describe the raytracing simulations. We present our results in 
Section~\ref{sect:results}. Finally we conclude in Section~\ref{sect:conclusions}.

\section{Coupled dark energy models}\label{sect:DEmodels}
In this work we consider weak gravitational lensing in the framework of coupled dark energy models. 
Dark energy is represented by a classical scalar field $\phi$ that evolves in a self-interaction potential $V(\phi)$ 
and interacts directly with cold dark matter particles by exchanging energy-momentum. This is due to a source term at 
the level of the background continuity equations of the Dark Energy and CDM components, characterised by a coupling 
function $\beta(\phi)$.

More quantitatively, the background dynamics for radiation (subscript $r$), baryons (subscript $b$), cold dark matter 
(subscript $c$) and Dark Energy scalar field (subscript $\phi$), are respectively described by the following set of 
equations:
\begin{eqnarray}
 \dot{\rho}_r+4H\rho_r & = & 0\;,\label{eqn:r} \\
 \dot{\rho}_b+3H\rho_b & = & 0\;,\label{eqn:b} \\
 \dot{\rho}_c+3H\rho_c & = & -\sqrt{\frac{2}{3}}\beta_c(\phi)\frac{\rho_c\dot{\phi}}{M_{Pl}}\;,\label{eqn:c}\\
 \ddot{\phi}+3H\dot{\phi}+V^{\prime}(\phi) & = & \sqrt{\frac{2}{3}}\beta_c(\phi)\frac{\rho_c}{M_{Pl}}\;,
 \label{eqn:phi}
\end{eqnarray}
where the Hubble function is given as usual by
\begin{equation}
 H^2=\frac{8\pi G}{3}\left(\rho_{\rm r}+\rho_{\rm c}+\rho_{\rm b}+\rho_\phi\right)\;,\label{eqn:H}
\end{equation}
and $M^2_{Pl}\equiv 1/8\pi G$ is the reduced Planck mass. The scalar field $\phi$ is expressed in units of 
$M_{\rm Pl}$, the overdot represents a derivative with respect to proper time and a prime stands for the derivative 
with respect to the scalar field.

The source terms in Eqs.~\ref{eqn:c}-\ref{eqn:phi} define the interaction between the dark matter and the dark energy 
components. The coupling function $\beta_c(\phi)$ controls the strength of the interaction and the sign of the term  
$\dot{\phi}\beta_c(\phi)$ controls the direction of the energy-momentum flow between the two coupled components, with 
a positive sign implying the transfer of energy-momentum from CDM to DE. The presence of the coupling term implies 
that the mass of the dark matter particles is not constant any more, but changes in time according to the following 
equation:
\begin{equation}
 \frac{d\ln(m_{\rm c}/M_{\rm Pl})}{dt}=-\sqrt{\frac{2}{3}}\beta_c\dot{\phi}\;.
\end{equation}
The sign of $\dot{m_c}$ depends therefore on the sign of the flow: a positive (negative) value of 
$\dot{\phi}\beta_{\rm c}(\phi)$ implies a decrement (increment) of the mass of dark matter particles. The equation of 
state of the dark energy component is given by $w_{\phi}\equiv P_\phi/\rho_\phi$, where the pressure $P_{\phi}$ and 
the density $\rho_{\phi}$ of the scalar field are defined as $P_{\phi}\equiv\dot{\phi}^2/2-V(\phi)$ and 
$\rho_{\phi}\equiv\dot{\phi}^2/2+V(\phi)$, respectively.

Coupled Dark Energy models do not affect only the background expansion history of the universe, but also the evolution 
of matter density perturbations due to the appearance of a long-range {\it fifth}-force term in the Euler equation. 
At the linear level, in the Newtonian limit and on sub-horizon scales, the linear perturbed equations read
\citep{Amendola2004,Pettorino2008,Baldi2011c}:
\begin{eqnarray}
 \ddot{\delta}_{\rm c} & = & -2H\left[1-\beta_c\frac{\dot{\phi}}{\sqrt{6}H}\right]\dot{\delta}_{\rm c}+4\pi
G[\bar{\rho}_{\rm b}\delta_{\rm b}+\bar{\rho}_{\rm c}\delta_{\rm c}\Gamma_{\rm c}]\;,\\
 \ddot{\delta}_{\rm b} & = & -2H\dot{\delta}_{\rm b}+
 4\pi G[\bar{\rho}_{\rm b}\delta_{\rm b}+\bar{\rho}_{\rm c}\delta_{\rm c}]\;.
\end{eqnarray}
In the previous equations, $\bar{\rho}_k$ represents the background density of the fluid $k$ and 
$\delta_k\equiv\delta\rho_k/\bar{\rho}_k$ its density perturbation. Note the presence of the factor 
$\Gamma_c\equiv 1+4\beta^2_c/3$ due to the presence of the fifth-force appearing only in the CDM equation. The term 
$\beta_c\dot{\phi}$ -- also appearing only in the CDM equation -- arises as a consequence of momentum conservation and 
effectively describes an additional friction term.

At the non-linear level, the acceleration experienced by DM particles is characterised by the two additional terms in 
the following equation:
\begin{equation}\label{eqn:friction}
 \dot{\vec{v}}_c=\beta_c\frac{\dot{\phi}}{\sqrt{6}}\vec{v}_c-
 \vec{\nabla}\left[\sum_c\frac{GM_c(\phi)\Gamma_c}{r_c}+\sum_b\frac{GM_b}{r_b}\right]\;,
\end{equation}
where $r_{\rm c,b}$ are the physical distances of the target coupled particle from the other CDM and baryonic 
particles, respectively. Effects of the friction term have been studied in the literature 
\citep{Amendola2004,Baldi2011c,Baldi2012b}.

\section{The CoDECS simulations}\label{sect:Nbody}
The basis for our lensing study is the suite of CoDECS N-body simulations \citep{Baldi2012}. Here we briefly describe 
them and we refer to \cite{Baldi2012} for a more in-depth discussion.

The CoDECS simulations are the largest suite of coupled dark energy simulations to date and are performed with a 
modified version \citep{Baldi2010} of the widely used TreePM N-body code {\small{GADGET}} \citep{Springel2005}. 
The code self-consistently simulates the evolution of structure formation in coupled dark energy models, taking into 
account the modified expansion history, the rise of a fifth-force and additional friction on each particle and the 
time variation of the dark matter particle mass.

The set of CoDECS simulations consists of two different types of runs, the L-CoDECS and the H-CoDECS runs. 
The H-CoDECS simulations are adiabatic hydrodynamical simulations of a box of only 80~Mpc/h comoving describing the 
evolution of an equal number of dark matter and gas particles ($512^3$). As our focus is on larger scales, we 
instead exploit the L-CoDECS runs, which follow the evolution of $1024^3$ DM particles and as many baryons in a box of 
comoving side of 1~Gpc/h. Both DM and gas particles are treated as collisionless particles, but they experience 
different dynamics, as a consequence of the interaction between the cold dark matter and the dark energy fluid. 
In fact, not properly taking into account the effect of the uncoupled baryonic fraction in interacting dark energy 
models would result in an incorrect evolution of structure formation. The run has a gravitational softening 
$\epsilon_{\rm s}=20$~kpc/h comoving; DM and baryon particles have a mass 
$m_{\rm DM}(z=0)=5.84\times 10^{10}~M_{\odot}/h$ and $m_{\rm b}=1.17\times 10^{10}~M_{\odot}/h$, respectively.

Six different cosmological models are simulated. The reference model is the standard $\Lambda$CDM model; three coupled 
dark energy models (EXP001, EXP002 and EXP003) are characterised by a constant positive coupling $\beta_c>0$ and an 
exponential self-interaction potential of the form $V(\phi)=A\exp{(-\alpha\phi)}$. 
Another model (EXP008e3) has the same potential but an exponential coupling, $\beta_c(\phi)=\beta_0\exp{(\beta_1\phi)}$ 
and finally the last model (SUGRA003) has a constant negative coupling, $\beta_c<0$ and a SUGRA \citep{Brax1999} 
self-interaction potential $V(\phi)=A\phi^{-\alpha}\exp{(-\phi^2/2)}$. 
We refer to Table~2 in \cite{Baldi2012} for values of the potential parameters in each case.

The normalization of the models is consistent with the WMAP7 cosmology \citep{Komatsu2011} and the linear matter power 
spectrum used to create initial conditions was computed with the publicly available code CAMB\footnote{www.camb.info} 
\citep{Lewis2000}. 
All the models have the same amplitude of perturbations at $z=z_{\rm CMB}$. 
Initial conditions for the simulations have been created starting from a glass distribution 
\citep{White1994,Baugh1995} evaluating particle displacements at $z=99$ using Zel'dovich approximation.

\section{Lensing and the raytracing simulations}\label{sect:lensing}
\subsection{The lensing observables}\label{subsect:lensing}
Due to the gravitational effects of matter on photons, light rays are deflected from their otherwise straight paths. 
The coherence scale of structures is negligible with respect to the cosmological distances involved in weak lensing 
studies, so it is reasonable to slice the matter distribution into thin shells using the well known {\it thin-lens} 
approximation. 
Under this hypothesis, cosmic lenses are effectively considered as two-dimensional objects whose projected mass 
distribution $\Sigma(\vec{\theta})$ on the lens plane is given by
\begin{equation}
 \Sigma(\vec{\theta})=\int\rho(\vec{\theta},l)dl\;,
\end{equation}
where $\vec{\theta}$ is the angular position on the lens plane and $l$ represents the direction along the line of 
sight.

The {\em convergence} is defined as
\begin{equation}\label{eqn:kappa}
 \kappa(\vec{\theta})\equiv \frac{\Sigma(\vec{\theta})}{\Sigma_{\rm crit}}\;,
\end{equation}
where $\Sigma_{\rm crit}$ represents the {\em critical surface density} and is defined as
\begin{equation}
 \Sigma_{\rm crit}\equiv\frac{c^2}{4\pi G}\frac{D_{\rm ds}}{D_{\rm d}D_{\rm s}}\;,
\end{equation}
where $D_{\rm ds}$, $D_{\rm d}$ and $D_{\rm s}$ are the angular-diameter distances between the lens and 
the source, between the observer and the lens and between the observer
and the source, respectively. 
The ratio of the distances represents the {\it lensing efficiency} and its maximum value is for lensing approximately 
half way between the observer and the source.

Under the thin-shell approximation, the lens is fully described by its convergence and therefore through the 
two-dimensional Poisson equation by the {\em lensing potential} $\Psi$
\begin{equation}
 \nabla^2_{\vec{\theta}}\Psi=2\kappa(\vec{\theta})\;,
\end{equation}
where the Laplacian is taken with respect to the angular position on the lens plane. The effect of the underlying 
matter distribution is to deflect the paths of light-rays and it is possible to show that the bending angle 
$\hat{\alpha}$ is related to the lensing potential $\Psi$ through:
\begin{equation}
 \hat{\alpha}=\nabla_{\vec{\theta}}\Psi\;.
\end{equation}

As a consequence of the light deflection, the observed image of the sources gets distorted. 
The mapping between the original source shape and the actual observed image, up to second order, is given by 
\cite[][]{Goldberg2005,Bacon2006}:
\begin{equation}
 \theta^{\prime}_i\simeq A_{ij}\theta_j+\frac{1}{2}D_{ijk}\theta_j\theta_k\;.
\end{equation}
In the previous equation, $A_{ij}\equiv\partial_j\theta^{\prime}_i$ is the Jacobian matrix of the mapping between the 
lensed and unlensed images, $\theta^{\prime}_i$ is the unlensed coordinate and the tensor $D_{ijk}$ describing the 
mapping at second order is the derivative of the Jacobian matrix with respect to the lensed coordinates $\theta_i$:
$D_{ijk}\equiv\partial_kA_{ij}$. In the previous equations, $\partial_k\equiv\partial/\partial\theta_k$.

While the convergence $\kappa$ gives a measure of the lensing strength weighted by the lens mass and the lensing 
efficiency, the distortions induced by gravitational lensing are quantified by the complex shear 
$\gamma=\gamma_1+i\gamma_2$, which is related to the second derivatives of the lensing potential $\Psi$ 
\cite[][]{Bartelmann2001}:
\begin{eqnarray}\label{eqn:gamma}
 \gamma_1 & = & \frac{1}{2}(\partial^2_1-\partial^2_2)\Psi\;,\\
 \gamma_2 & = & \partial^2_{12}\Psi\;.
\end{eqnarray}
The elements of the matrices $A_{ij}$ and $D_{ijk}$ are conveniently expressed as a function of the convergence, and 
of the shear components and its derivatives \cite[see e.g.][]{Goldberg2005,Bacon2006,Pace2011}.
A suitable combination of the derivatives of the shear components gives rise to two new quantities, the 1- 
and 3-flexion ($F$ and $G$ respectively) \cite[][]{Bacon2006}:
\begin{eqnarray}\label{eqn:flexion}
 F & \equiv & F_1+iF_2 = (\gamma_{1,1}+\gamma_{2,2})+i(\gamma_{2,1}-\gamma_{1,2})\;,\\
 G & \equiv & G_1+iG_2 = (\gamma_{1,1}-\gamma_{2,2})+i(\gamma_{2,1}+\gamma_{1,2})\;.
\end{eqnarray}

The results derived so far are valid only in the case of a single lens between the observer and the source; however 
the whole formalism can be generalised to the case of a continuous matter distribution. 
The procedure is very similar to the case of the single lens. The cosmic volume can be sliced in sufficiently small 
sub-volumes whose thickness along the line-of-sight is sufficiently small with respect to the distances involved 
(namely the distances between the observer, the lenses and the sources). 
Therefore the thin-lens approximation should be valid and the matter distribution can be projected on a plane, and 
consequently the lensing potential can be evaluated using the Poisson equation. 
Also for multiple lenses therefore all the information is embedded in the lensing potentials on the slices.  
The final quantities (convergence, shear and flexion) can now be estimated on the source plane as the weighted sum of 
the contributions from all the different lensing planes. 
Unlike the single lens case, the Jacobian matrix is no longer symmetric, due to the fact that rotation of the light 
bundles can occur. 
As shown with the help of numerical simulations \cite[see][]{Jain2000}, the rotation term is very small and can be 
safely neglected; we verified that this is indeed the case for our simulations.

\subsection{Raytracing simulations}\label{sect:raytracing}
Raytracing techniques consist of shooting rays through an N-body simulation and evaluating the deflection angle and 
related quantities by taking into account the underlying matter distribution.

The light cones are constructed by stacking snapshots of a single simulation; the snapshots are placed along the line 
of sight so that the light cone distance to the centre of the simulation corresponds with the time of the snapshot.  
We put sources at $z_{\rm s}=1$ ($z_{\rm s}=2$) using 10 (13) snapshots.
Sufficient snapshots are stacked in order to avoid gaps in the matter distribution; this leads to some overlap in the 
stacking, which we account for by including only the volumes that do not overlap to the following snapshot in the 
stack.

Since each snapshot represents the same matter distribution at different cosmic times, the structures will be at 
roughly the same position in each volume. 
To avoid artificial correlations between the matter density at different redshifts, we coherently rotate and shift 
particle positions in each snapshot by a random amount, taking advantage of the periodic boundary conditions, so that 
particles leaving the simulated box on one side, re-enter on the opposite one. 
Finally, in order to estimate the sample variance errors, 100 different realizations were created for each model by 
using different shifts and rotations of the snapshots.

\begin{table*}
 \centering
 \caption[]{Characteristic parameters for the raytracing simulations and normalization of the power spectrum.}
 \begin{tabular}{lcccc}
  \hline
  \hline
  Models & Opening angle (degrees) & Resolution (arc sec) & Source
  comoving distance for $z_{\rm s}=1$ ($Mpc/h$) & $\sigma_8(z=0)$\\
  \hline
  $\Lambda$CDM & 24.34 & 21.39 & 2355.14 & 0.809 \\
  EXP001       & 24.36 & 21.41 & 2353.19 & 0.825 \\
  EXP002       & 24.39 & 21.44 & 2350.43 & 0.875 \\
  EXP003       & 24.43 & 21.47 & 2346.42 & 0.967 \\
  EXP008e3     & 24.55 & 21.58 & 2335.29 & 0.895 \\
  SUGRA003     & 25.23 & 22.17 & 2272.19 & 0.806 \\
  \hline
  \label{tab:rayt_params}
 \end{tabular}
 \begin{flushleft}
  \vspace{-0.5cm}
  {\small}
 \end{flushleft}
\end{table*}

The opening angle of the raytracing simulation is evaluated with the comoving size and distance of the source plane. 
Due to a different background evolution, the distance of the source from the observer will slightly change with the 
model. For the reference $\Lambda$CDM model, the opening angle is $\theta=24.34$ ($\theta=15.4$) degrees on a side 
and the resolution of the map is 21.39 (13.5) arcsec for sources at $z_{\rm s}=1$ ($z_{\rm s}=2$). In 
Table~\ref{tab:rayt_params} we report these parameters for the other models considered in this paper. Although very 
similar, the map resolution is higher for the $\Lambda$CDM run since the sources are further away from the observer 
with respect to the other models. As seen in the second column of Table~\ref{tab:rayt_params}, the opening angle is 
minimum for the $\Lambda$CDM run and maximum for the SUGRA003 model. This is due to the fact that the opening angle 
is evaluated as the ratio of the comoving size of the source plane and its comoving distance with respect to the 
observer. (The comoving box size is the same for all the different models studied.) In the last column of 
Table~\ref{tab:rayt_params} we present the normalization of the matter power spectrum for the different models. Note 
how the models have considerably different values of $\sigma_8$. In particular all the EXP models have a higher 
normalisation with respect to the $\Lambda$CDM model, while for the SUGRA model it is approximately the same. As we 
will see later, this will have a crucial importance in explaining the differences between the models. All the models 
have the same amplitude of scalar perturbations at $z_{\rm CMB}$, therefore the different normalizations at $z=0$ 
reflect the different structure evolution, as shown by the growth factor in Fig.~2 of \cite{Baldi2012}. All the EXP 
models show a monotonic increase of the ratio of the linear growth factor with respect to the $\Lambda$CDM model, 
while the SUGRA model recovers the amplitude of perturbations at $z=0$.

We next briefly sketch how we created the lensing maps. For more details we refer the reader to 
\cite{Hamana2001,Pace2007.1,Pace2011}. For each lens plane (which corresponds roughly to the snapshots), we evaluate 
the projected matter density field and then with FFT techniques we can recover the corresponding lensing potential. 
The selected particles are projected parallel to the line-of-sight on a two dimensional grid of $4096^2$ pixels; to 
assign particles to pixels, we use the triangular-shape-cloud (TSC) method as outlined in \citet{Hockney1988}.
Following the prescription of \citet{Hamana2001}, we project particles over a regular grid to obtain the projected 
overdensity field for each lens plane:
\begin{equation}
 \delta^{{\rm proj},k}_{ij}=\frac{M^{k}_{ij}}{A_{k}\bar{\rho}_k}-L_k\;,
\end{equation}
where the index $k$ runs over the lens planes, $M^{k}_{ij}$ is the mass projected in the box $k$ on the pixel $(ij)$, 
$A_k$ the pixel area and $L_k$ the size of the projected box (in our case it will be smaller than the full box size, 
due to the overlapping volumes). Finally, $\bar{\rho}_k$ is the comoving background density. Note that 
$\bar{\rho}(a)=\bar{\rho}_0\Omega_{\rm 0}$ only for the $\Lambda$CDM model, since for other models the time evolution 
of the matter density parameter is different from the standard $(1+z)^3$ behaviour.

Formally the lensing potential is still evaluated via the two-dimensional Poisson equation, but in this case we also 
have to take into account the different matter density parameter evolution. We therefore write the Poisson equation as 
\cite[][]{Hamana2001}
\begin{equation}\label{eqn:Poisson}
 \nabla^2_{\vec{x}}\Psi^k=\frac{8\pi G\bar{\rho}_k}{c^2}\delta^{{\rm proj},k}\;.
\end{equation}
Eq.~\ref{eqn:Poisson} can be solved via FFT techniques taking advantage of the periodic boundary conditions. 
As shown above (see Sect.~\ref{subsect:lensing}), the lensing potential fully characterises our system. 
We can therefore obtain all the lensing quantities we are interested in via standard finite difference techniques.

The raytracing simulations are based on stacking multiple-lensing planes and the result evaluated on the source plane 
is given by adding the weighted contribution of all the planes between the source and the observer. 
Suppose the cosmic volume is sliced into $N$ lens planes and the source plane is labelled as $N+1$. 
Light rays are shot from the observer and create a regular grid on the first lens plane. 
The bend angle on a given plane $k$ is related to the image position $\vec{\theta}_1$ on the first lens plane ($N=1$) 
through the relation
\begin{equation}\label{eqn:theta}
 \vec{\theta}_k = \vec{\theta}_1-\sum_{i=1}^{k-1}\,\frac{f_K(w_k-w_i)}{f_K(w_k)a_i}\nabla_{\vec{x}}\Psi_i(\vec{x})\;,
\end{equation}
where $w$ is the comoving distance, $a_i$ the scale factor of the lens plane, $f_K$ a function depending on the 
cosmology and $\Psi_i(\vec{x})$ the Newtonian potential projected along the line-of-sight on each lens plane.

The Jacobian on each lens plane is obtained by differentiating Eq.~\ref{eqn:theta} with respect to $\vec{\theta}_1$. 
By defining $A_k\equiv\partial\vec{\theta}_k/\partial\vec{\theta}_1$ and indicating with $U_k$ the matrix whose 
elements are the second derivatives of the lensing potential we derive the following equation:
\begin{equation}\label{eqn:Jacobian}
 A_k=I-\sum_{i=1}^{k-1}\frac{f_K(w_i)f_K(w_k-w_i)}{f_K(w_k)a_i}U_iA_i\;,
\end{equation}
where $I$ represents the identity matrix. 
A further derivative of Eq.~\ref{eqn:Jacobian} with respect to $\vec{\theta}_i$ gives a similar recursive relation for 
the two flexions
\cite[][]{Pace2007.1,Pace2011}:
\begin{equation}\label{eqn:Dmatrix}
 D_k^{1,2}=-\sum_{i=1}^{k-1}\frac{f_K(w_i)f_K(w_k-w_i)}{f_K(w_k)a_i}[f_K{w_i}G_U^{1,2}+U_iD_i^{1,2}]\;.
\end{equation}
In the previous equation, $G_U\equiv\nabla_{\vec{x}}U$ is a tensor containing the third derivatives of the lensing 
potential.

\begin{figure*}
 \centering
 \includegraphics[width=0.42\textwidth]{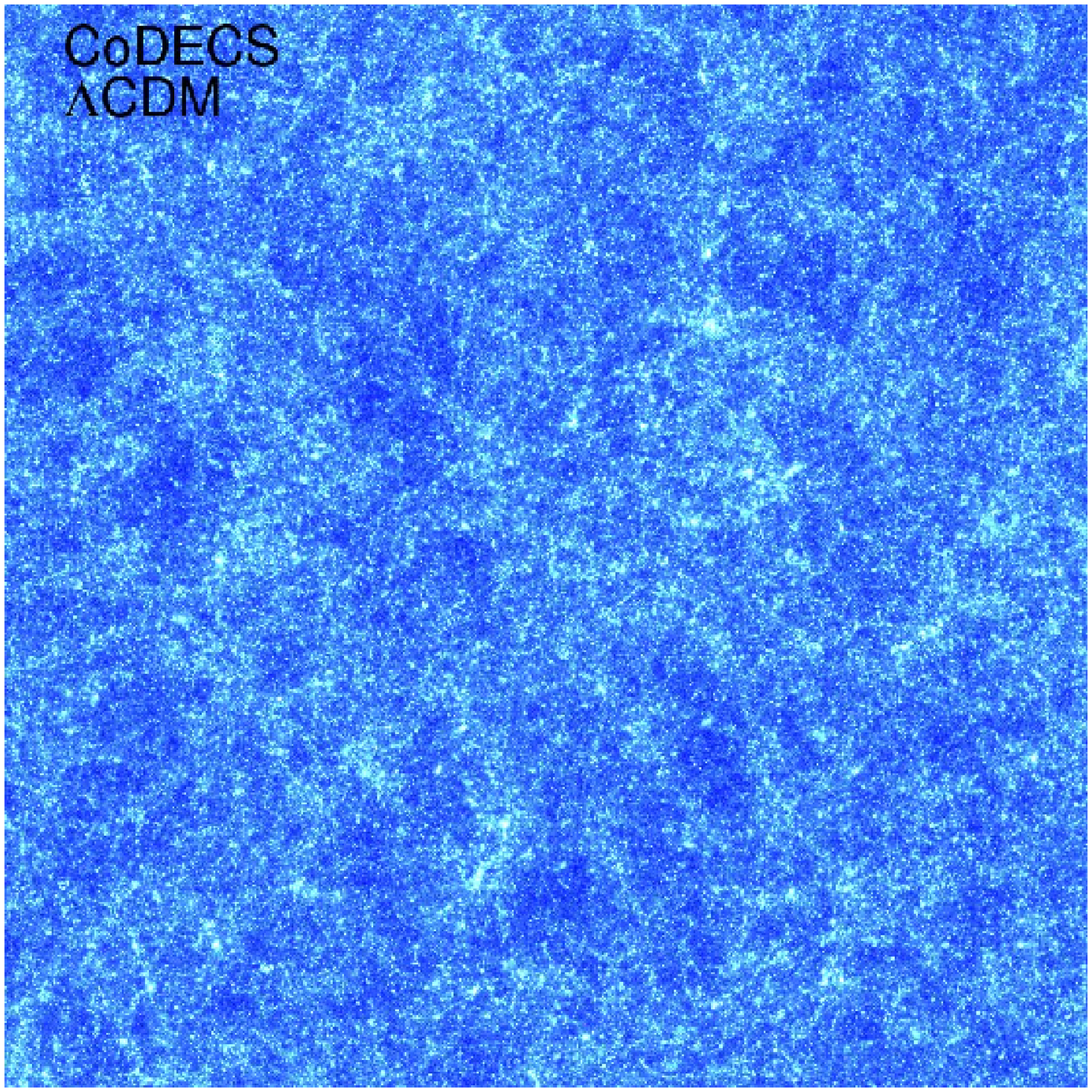}
 \includegraphics[width=0.42\textwidth]{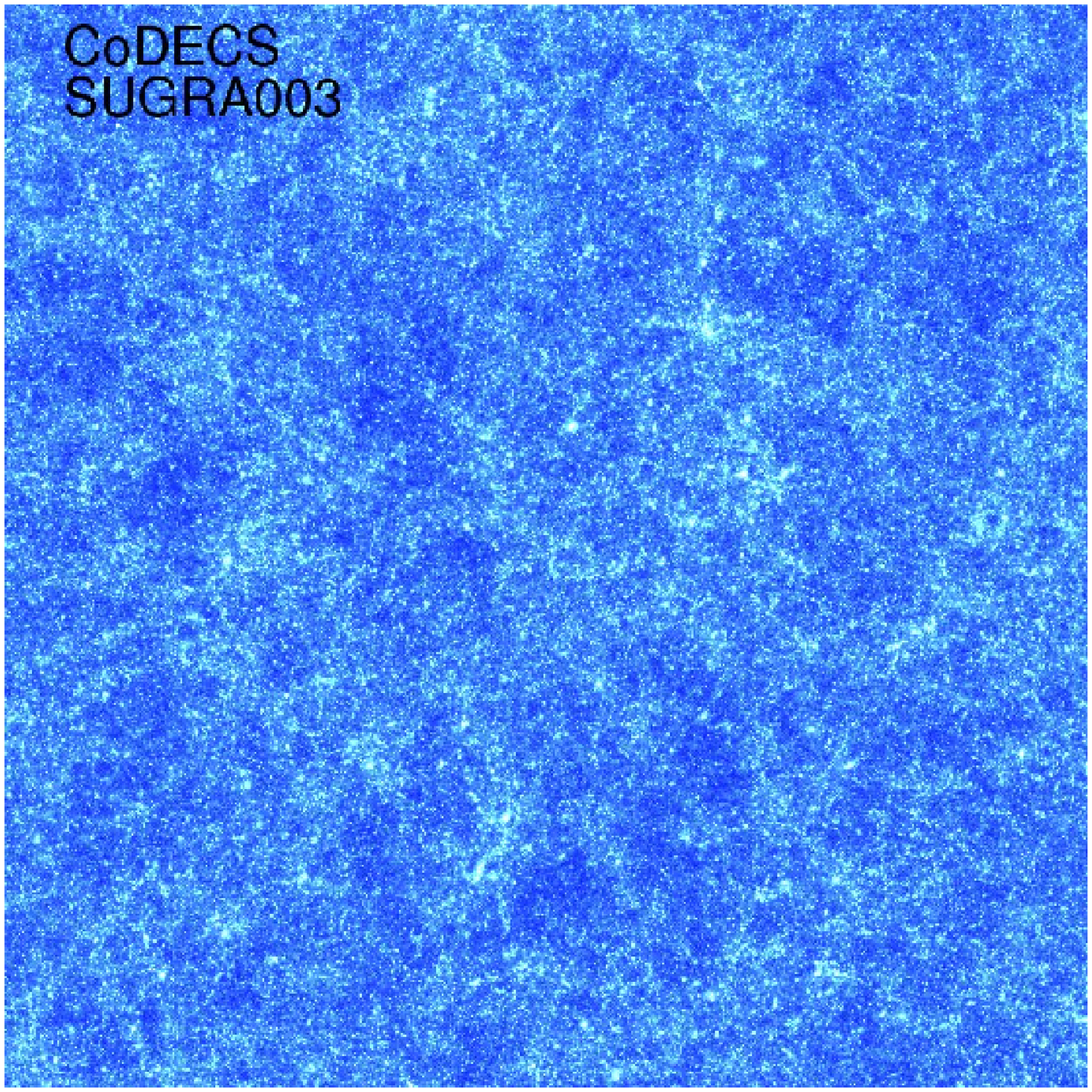}
 \includegraphics[width=0.42\textwidth]{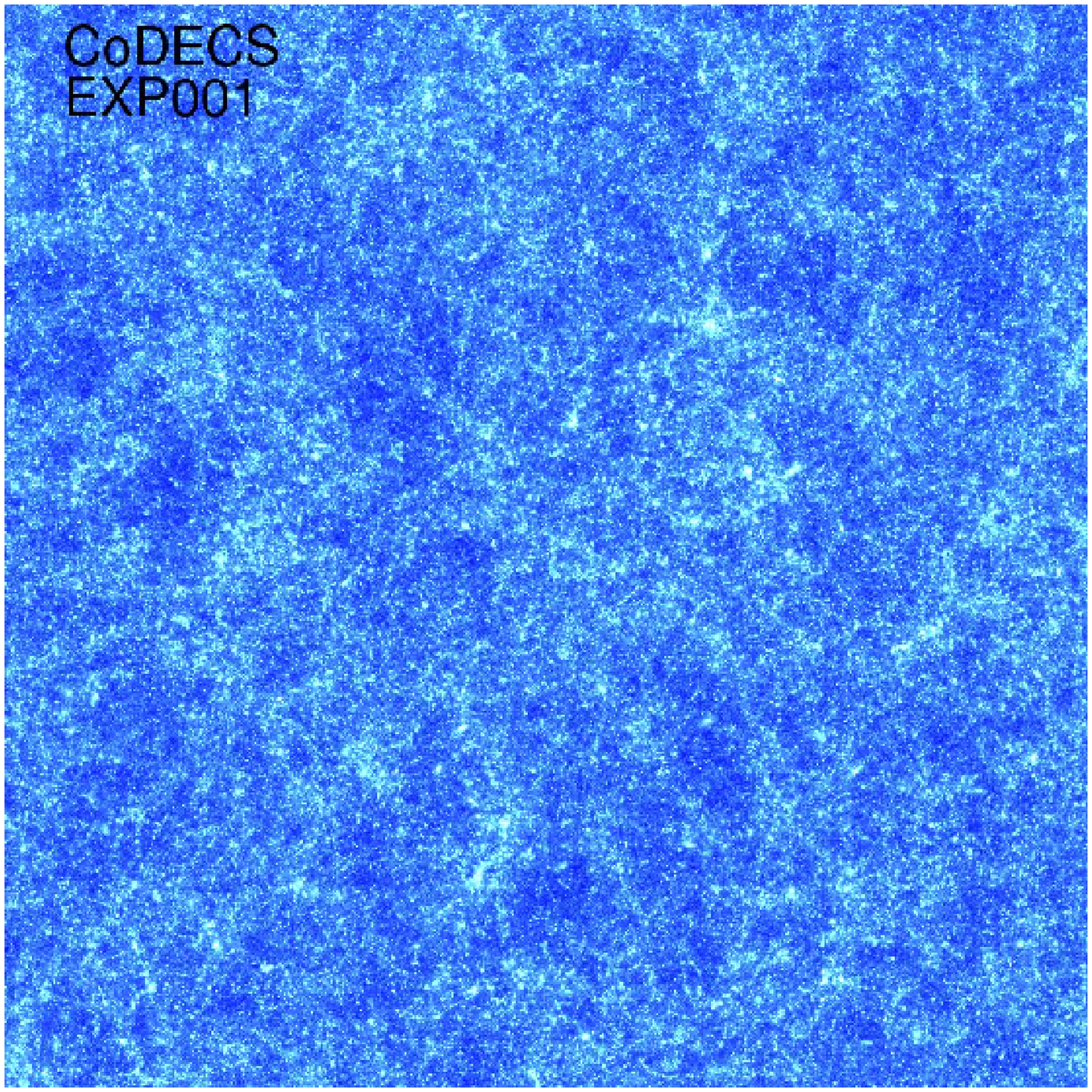}
 \includegraphics[width=0.42\textwidth]{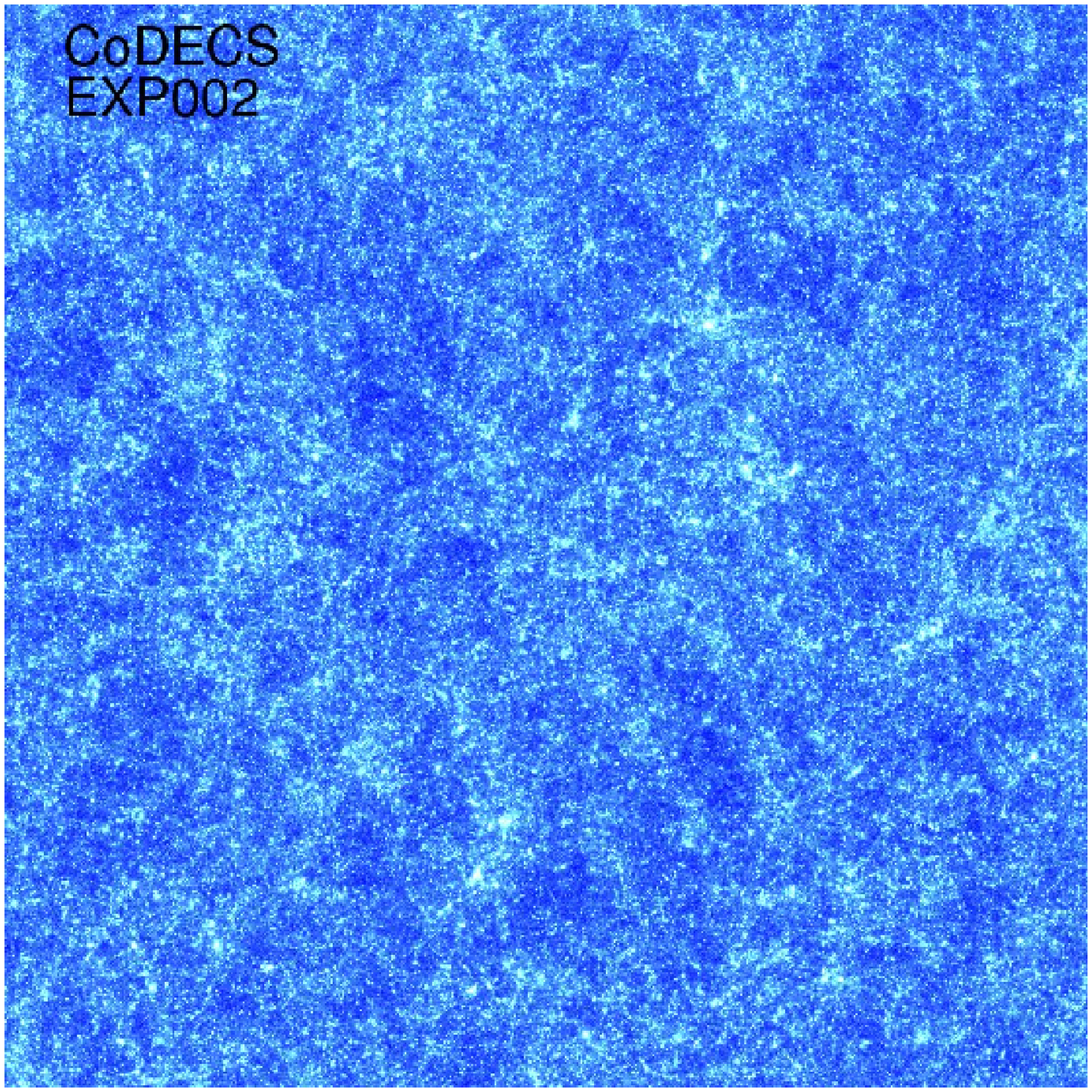}
 \includegraphics[width=0.42\textwidth]{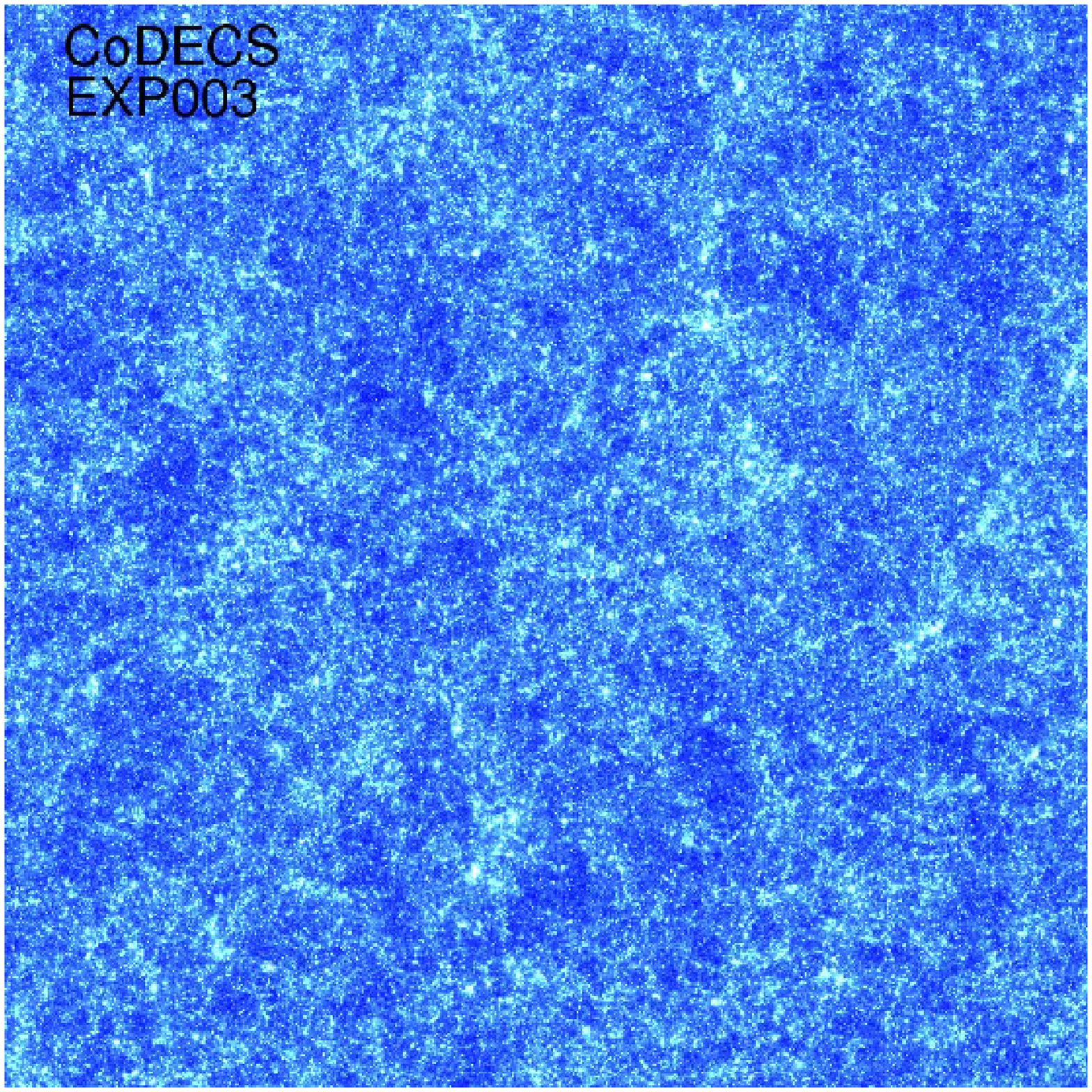}
 \includegraphics[width=0.42\textwidth]{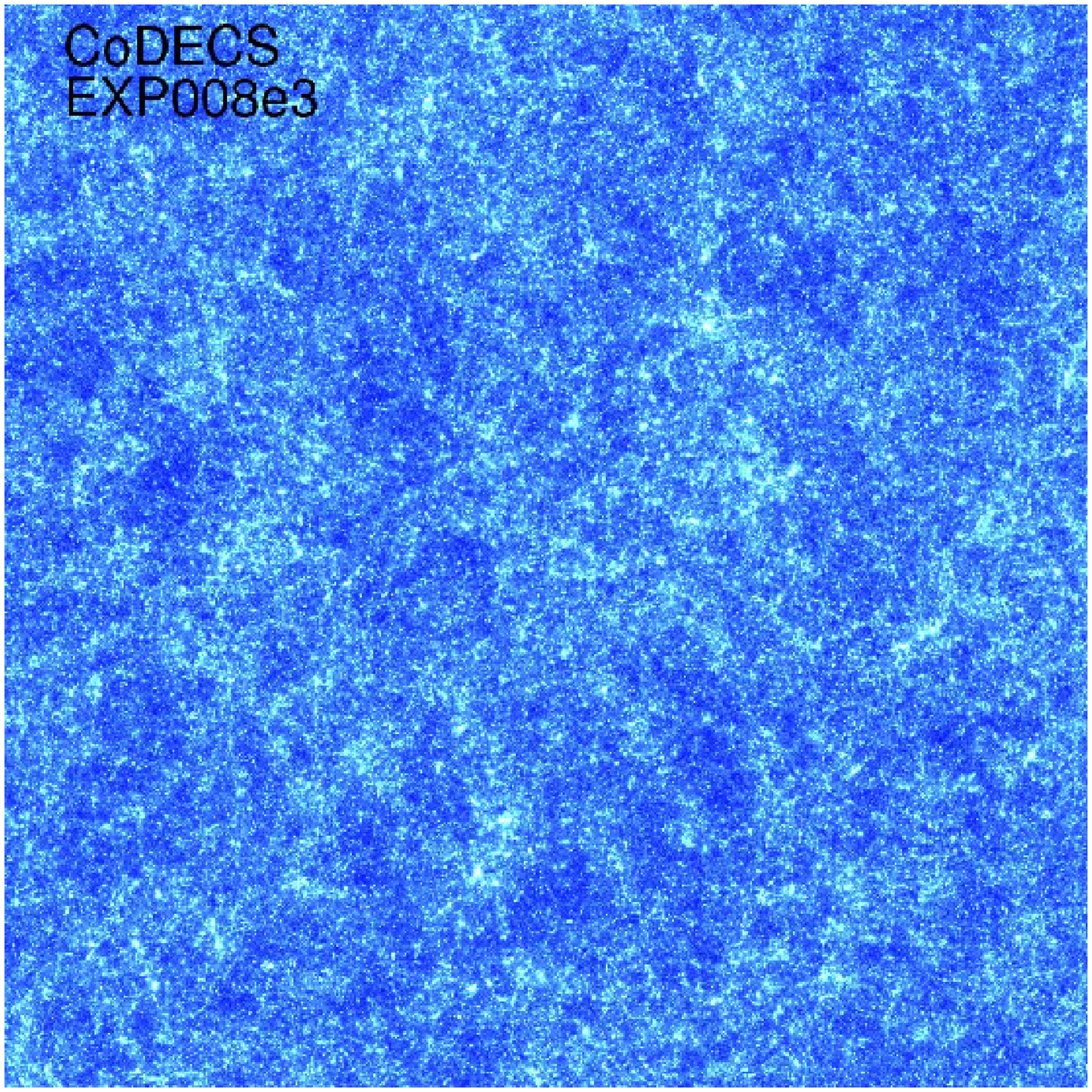}
 \caption{The effective convergence for one particular realization of the raytracing simulations used in this work. 
Sources are at $z_{\rm s}=1$. Colour range is the same for all the models. Different panels refer to different 
cosmological models, as labelled. We refer to Table~\ref{tab:rayt_params} for the field of view spanned by each 
simulation.}
 \label{fig:kappa_maps}
\end{figure*}

On the source plane, the matrices $A$ and $D^{1,2}$ can be related to observable quantities like the effective 
convergence, shear and derivatives of the shear (that combined together provide the 1- and 3-flexion). They read
\begin{eqnarray}\label{eqn:matrixes}
 A_{N+1} & = & \left( \begin{array}{cc}
    1-\kappa-\gamma_1 & -\gamma_2+\omega \\
    -\gamma_2-\omega  & 1-\kappa+\gamma_1
    \end{array}\right)\nonumber\\
 D^1_{N+1} & = & \left( \begin{array}{cc}
    -2\gamma_{1,1}-\gamma_{2,2} & -\gamma_{2,1}+\omega_1 \\
    -\gamma_{2,1}+\omega_2      & -\gamma_{2,2}+\omega_3 \\
  \end{array} \right)\\
 D^2_{N+1} & = & \left( \begin{array}{cc}
    -\gamma_{2,1}+\omega_4 & -\gamma_{2,2}+\omega_5 \\
    -\gamma_{2,2}+\omega_6 & 2\gamma_{1,2}-\gamma_{2,1} \\
  \end{array} \right) \nonumber\;.
\end{eqnarray}
The scalar $\omega$ is called the rotation term and describes the rotation of the light bundle due to multiple 
lenses. Following \cite{Bacon2009}, we identify the additional quantities $\omega_i$, with $i=1-6$, as a combination 
of the components of the twist $C\equiv C_1+\imath C_2$ and the turn $T=T_1+\imath T_2$:
\begin{eqnarray}
\omega_1 & = & -\frac{1}{2}(C_1+T_1+T_2) \\
\omega_2 & = & -\frac{1}{2}(C_1+T_1-T_2) \\
\omega_3 & = & T_1+T_2 \\
\omega_4 & = & T_1-T_2 \\
\omega_5 & = & \frac{1}{2}(C_1-T_1-T_2)\\ 
\omega_6 & = & -\frac{1}{2}(C_2+T_1+T_2)\;.
\end{eqnarray}
We performed several tests on our raytracing simulations and showed that our results are largely unaffected by their 
presence, so we neglect them below.

As an example, in Fig.~\ref{fig:kappa_maps} we show one realization of the effective convergence maps. For all the 
models we used the same random seed so to have the same distribution of structures along the light-cone. As expected, 
the main pattern of the effective convergence $\kappa$ is very similar for all the models, but some differences can 
be noticed even by eye. In particular we observe that the realization for the SUGRA003 model, despite having 
basically the same normalization of the $\Lambda$CDM model, shows less pronounced structures and lower convergence 
peaks. The EXP003 model, on the other hand, presents a larger number of structures and higher peaks, due to the 
higher normalization of the matter power spectrum. Similar conclusions can be drawn for the other models, where 
differences becomes more evident when the matter power spectrum normalization increases.

In the following sections these differences, already visible by eye, will be analysed in a more quantitative way 
using various statistical tests and will be explained in terms of the different evolution of the matter density 
perturbations in the various cosmological models.

\section{Results}\label{sect:results}
In this section we describe the results we obtained from the analysis of our simulated maps. In Sect.~\ref{sect:PS} 
we discuss results related to the lensing power spectrum and in Sect.~\ref{sect:shear_aperture} 
and~\ref{sect:CF} the shear in aperture and the correlation function, respectively. 
In Sect.~\ref{sect:PDF} we illustrate results regarding the probability distribution function (PDF) of some of the 
lensing quantities; in Sect.~\ref{sect:high_order} we examine higher order moments such as the variance, the skewness 
and the kurtosis. All results shown in this section are the average (or median) values computed over 100 different 
realizations, while the error bars (shown only for the reference $\Lambda$CDM model for clarity reasons) represent 
the r.m.s. (or quartiles) of the same 100 realizations.

\subsection{Power spectrum}\label{sect:PS}
We begin with the study of the power spectra of different lensing observables in the simulated maps. The shear (or 
effective convergence) power spectrum is a very important observational quantity that can be used to probe the 
underlying cosmological model, to infer the normalization of the matter power spectrum and the growth of structures. 
In the Born approximation, the shear power spectrum is related to the integral along the line of sight of the matter 
power spectrum, weighted by distance factors taking into account the geometry of the system (in particular the 
relative distances between source, lens and observer). As explained before, in cDE models, dark matter evolution no 
longer follows the $a^{-3}$ time evolution, and the time evolution of the power spectrum is affected by this. The 
relation between the matter $P_{\delta}(k)$ and the lensing $P_{\kappa}(\ell)$ power spectra is given by 
\citep[see also][]{Beynon2012}
\begin{equation}\label{eqn:Pl}
 P_{\kappa}(\ell)=\frac{9}{4}\left(\frac{H_0}{c}\right)^4\int_0^{\chi_H}d\chi W^2(\chi)f(a)
 P_{\delta}\left(k=\frac{\ell}{\chi},\chi\right)\;,
\end{equation}
where $f(a)=a^4\Omega^2_{m}(a)E^4(a)$, $E\equiv H/H_0$ is the dimensionless Hubble function, $\chi_H$ the comoving 
distance to the horizon and $\Omega_{\rm m}(a)$ the matter density evaluated at the scale factor $a$.

In the weak lensing regime (the one of interest for this work), the spectra of shear $\gamma$, reduced shear $g$, 
convergence $\kappa$ and flexions $F$ and $G$ are all inter-related; in particular
\begin{eqnarray}
 & & P_{\gamma}(\ell) = P_{\kappa}(\ell)=P_{g}(\ell)\\
 & & P_{F}(\ell) = P_{G}(\ell)=\ell^2 P_{\kappa}(\ell)\;.
\end{eqnarray}
In reality the true observable is the power spectrum of the reduced shear $g$, defined as
\begin{equation}
 g\equiv \frac{\gamma}{1-\kappa}\;,
\end{equation}
and its spectrum is approximately the same as the one for cosmic shear, as distortions of the images are very small.

To evaluate the power spectrum of the different lensing quantities we consider each pixelated map and evaluate 
its Fourier transform on the grid. We then multiply each map in Fourier space by its complex conjugate and determine 
the corresponding frequency at each pixel. A further binning of the spectrum obtained in this way gives the final 
smoothed result.

\begin{figure}
 \centering
 \includegraphics[width=0.3\textwidth,angle=-90]{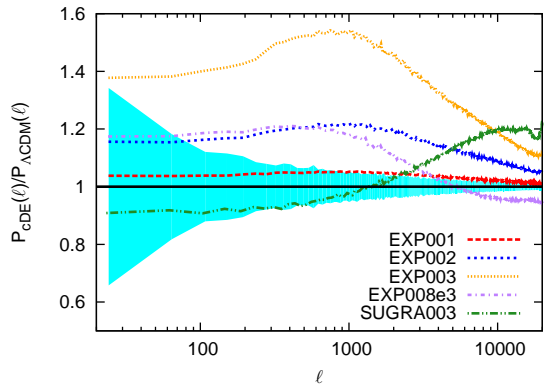}
 \caption[]{Ratio between the convergence power spectrum of the coupled dark energy models and the reference 
   $\Lambda$CDM model as a function of the multipole $\ell$. Different colours and line styles represent different 
   models. The $\Lambda$CDM model is shown with solid black line, the EXP001 model with dashed red line, the EXP002 
   model with blue short-dashed line, the EXP003 model with orange dotted line, the EXP008e3 model with the light 
   violet dashed-dotted line and finally the SUGRA003 model with the green dashed-dot-dotted line. The curves and the 
   shaded region (shown only for the $\Lambda$CDM model) represent the average and the r.m.s. obtained from 100
   different realizations, respectively.}
 \label{fig:PSkappa}
\end{figure}

In Fig.~\ref{fig:PSkappa} we show the ratio between the lensing power spectrum of the coupled dark energy models and 
the reference spectrum of the $\Lambda$CDM model. Since the power spectrum for the (reduced) shear is identical to 
that of the effective convergence and the spectra of the two flexions are simple functions of the convergence, we 
limit ourself to the ratios for the effective convergence. We show our results for wavelengths up to 
$\ell\approx 2\times 10^4$, since for higher values, the noise in our simulations starts dominating. 
The model EXP001 has a slightly different $\sigma_8$ and differences are of the order of 
few percent; it is well within the error bars at all the wavelengths probed in our raytracing simulations, making it 
very difficult to distinguish it from the reference model.
It is similarly difficult to discriminate between the SUGRA003 model and the $\Lambda$CDM model on large scales, 
since the ratio is well within the simulation uncertainty from cosmic variance. 
The largest deviations appear for the model EXP003 where on large scales the differences are around 40\%.

In the EXP models the power is higher than for the $\Lambda$CDM model: this is due to the faster growth of 
perturbations and therefore to higher matter power spectrum normalization; this is reflected directly in the 
different amplitude at small $\ell$. 
The SUGRA003 model is quite different. Despite having basically the same $\sigma_8$ normalization as the $\Lambda$CDM 
model, we notice approximately 10\% less power up to $\ell\approx 1000$. This is easily explained in terms of the 
evolution of the matter density parameter, which for the SUGRA003 model is consistently smaller than that of the 
$\Lambda$CDM model at the redshifts of interest for this work.  This is due to the evolution of the dark matter mass. 
As shown in Eq.~\ref{eqn:Pl}, the lensing power spectrum is proportional to the matter density parameter, therefore a 
deficit in this quantity will directly translate into a lower power spectrum. 
These conclusions follow closely and reproduce the results on the matter power spectrum presented in \cite{Baldi2012} 
also in the weak lensing regime.

\begin{figure}
 \centering
 \includegraphics[width=0.3\textwidth,angle=-90]{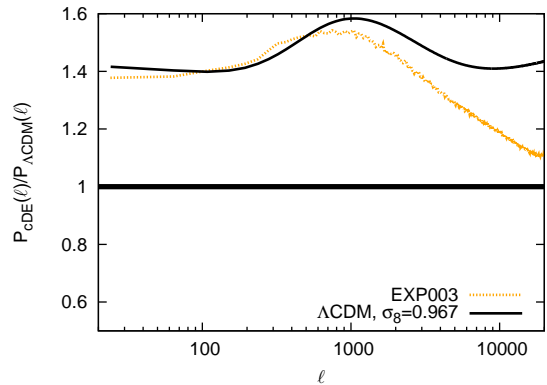}
 \caption[]{Ratio between the convergence power spectrum of the EXP003 (orange dotted line) and the reference 
$\Lambda$CDM model. For comparison we show the same ratio obtained analytically for a $\Lambda$CDM model having the 
same normalization $\sigma_8=0.967$ as the EXP003 model (black thin solid line).}
 \label{fig:LCDMps8}
\end{figure}

It is also interesting to notice an increase in the power with a peak at $\ell\approx 1000$, followed by a later 
decrease in the region dominated by the shot noise (an increase for the SUGRA003 model). 
These results, including the increase in the ratio, are in good qualitative agreement with \cite{Carbone2013}, though 
that work probed a much smaller range of multipoles than in this work. 
It is worth understanding whether the differences that arise are purely due to the different growth rate and power 
spectrum normalisation, or reflect deeper physical differences in the models. 
To address this question, we evaluate analytically the lensing power spectrum for a $\Lambda$CDM model with the same 
normalization $\sigma_8=0.967$ as the EXP003 model and we show our results in Fig.~\ref{fig:LCDMps8}. On large scales, 
we observe a fairly good agreement between the EXP003 model and the $\Lambda$CDM model with higher power spectrum 
normalization. 
(The lack of power for the largest modes is due to the missing power in the simulations arising from its finite size.) 
The increase of power we observe for the cDE model at higher multipoles also appears in this case, so it is evidently 
simply the result of the different normalization.  This is consistent with the different $\sigma_8$; although the 
amplitude is lower, it is in agreement with what was found for the three-dimensional matter power spectrum in 
\cite{Baldi2012}. 
The peak originates from the different evolution of the non-linear matter power spectrum; in particular, models 
with a higher normalization will have non-linear effects kicking in at lower $\ell$s with respect to a model with a 
lower normalisation. The feature occurs precisely at the linear-non-linear transition scale for the model with higher 
normalisation (at the redshift being probed.)

However, the normalisation is not the only effect at play; at higher $\ell$s, the EXP003 spectrum drops away from the 
analytical $\Lambda$CDM spectrum and the agreement is limited up to $\ell\approx 1000$. At the largest 
multipoles, we do not expect agreement between the analytic and simulations due to the finite resolution of the 
pixels in the ray-traced maps \citep[see also][]{Pace2007.1,Pace2011}. 
Comparing analytic and simulated $\Lambda$CDM spectra, these effects are seen to suppress the  spectra above 
$\ell\approx 4000$, and so below this any suppressions we see in the coupled dark energy models are believed to 
result from the modified physics. 
In order to examine the impacts at higher resolutions, we focus our attention on ratios of simulations where the 
pixel smoothing effects should cancel. 
Note that these smoothing effects equally impact the shear in aperture (Sect.~\ref{sect:shear_aperture}) and the 
probability distribution function (Sect.~\ref{sect:PDF}) observables.

While there is consensus that the reduced shear is the truly observable shear quantity, for flexion several 
possibilities have been considered. 
\cite{Viola2012} defined the reduced flexions in analogy to the reduced shear:
\begin{eqnarray}
 \mathcal{F} & = & \frac{F}{1-\kappa}\label{eqn:rFVM}\\
 \mathcal{G} & = & \frac{G}{1-\kappa}\label{eqn:rGVM}\;,
\end{eqnarray}
while \cite{Schneider2008} instead studied
\begin{eqnarray}
 \mathcal{F} & = & \frac{F+gF^{\ast}}{1-\kappa}\label{eqn:rFSR}\\
 \mathcal{G} & = & \frac{G+gF}{1-\kappa}\label{eqn:rGSR}\;,
\end{eqnarray}
where $F^{\ast}$ represents the complex conjugate of $F$. 
We have created reduced flexion maps according to both definitions and find that the results are very similar. 
In particular, ratios between the spectra of the reduced flexions for the coupled dark energy models and the 
$\Lambda$CDM model are equivalent to what is found for the unreduced flexion.

It is well known that it is possible to gain information on the time evolution of the large scale structure of the 
Universe by following a tomographic approach, i.e. studying the lensing effects produced on sources located at 
different redshifts. To investigate this issue in the context of coupled dark energy models, we have also used a set 
of effective convergence maps created for sources at $z_{\rm s}=2$, and we have evaluated for each model the ratio of 
the power spectra for sources at $z_{\rm s}=2$ and the ones for sources at $z_{\rm s}=1$. The aim is to see whether 
there is any signature due to the coupling that would make the ratio dependent on the multipoles in a peculiar way. 
We find that this is unfortunately not the case, since all the ratios are very similar to what is found for the 
$\Lambda$CDM model. Small differences are seen at very high $\ell$, where unfortunately we cannot completely trust 
our results due to the increase of noise and to resolution effects. Therefore the study of the convergence power 
spectrum with sources at different redshifts seems not to add any further information to what is inferred from the 
analysis at $z_{\rm s}=1$. This is due to the combination of the evolution of the dark matter parameter and friction 
terms.

We have seen that the ratios between the convergence power spectra of the different models faithfully reproduce 
the behaviour of the matter power spectrum, as explained in detail in \cite{Baldi2012}. In particular the EXP (SUGRA) 
models show a higher (lower) spectrum amplitude. 
There will also be degeneracies between the EXP and the $\Lambda$CDM model with respect to different values of 
$\sigma_8$ and between the SUGRA and the $\Lambda$CDM model with respect to different values for the matter density 
parameter $\Omega_{\rm m}$. To investigate these degeneracies, it would be necessary to run a larger suite of N-body 
simulations covering an array of models. This is beyond the scope of the present work, where we focus on the study of 
the effects of the coupling between dark matter and dark energy on the lensing observables.

\subsection{Shear in aperture}\label{sect:shear_aperture}
An alternative statistic to the power spectrum is the shear in aperture. 
The shear in aperture represents the variance of the shear field within a circular aperture of radius $\theta$ and it 
is related to the power spectrum by 
\begin{equation}\label{eqn:shear_aperture}
 \left|\gamma_{\rm av}(\theta)\right|^2\equiv 2\pi\int_0^{\infty}d\ell\ell
P_{\gamma}(\ell)\left[\frac{J_1(\ell\theta)}{\pi\ell\theta}\right]^2\;,
\end{equation}
where $J_1(x)$ is the first-order Bessel function of the first kind.

\begin{figure}
 \includegraphics[width=0.3\textwidth,angle=-90]{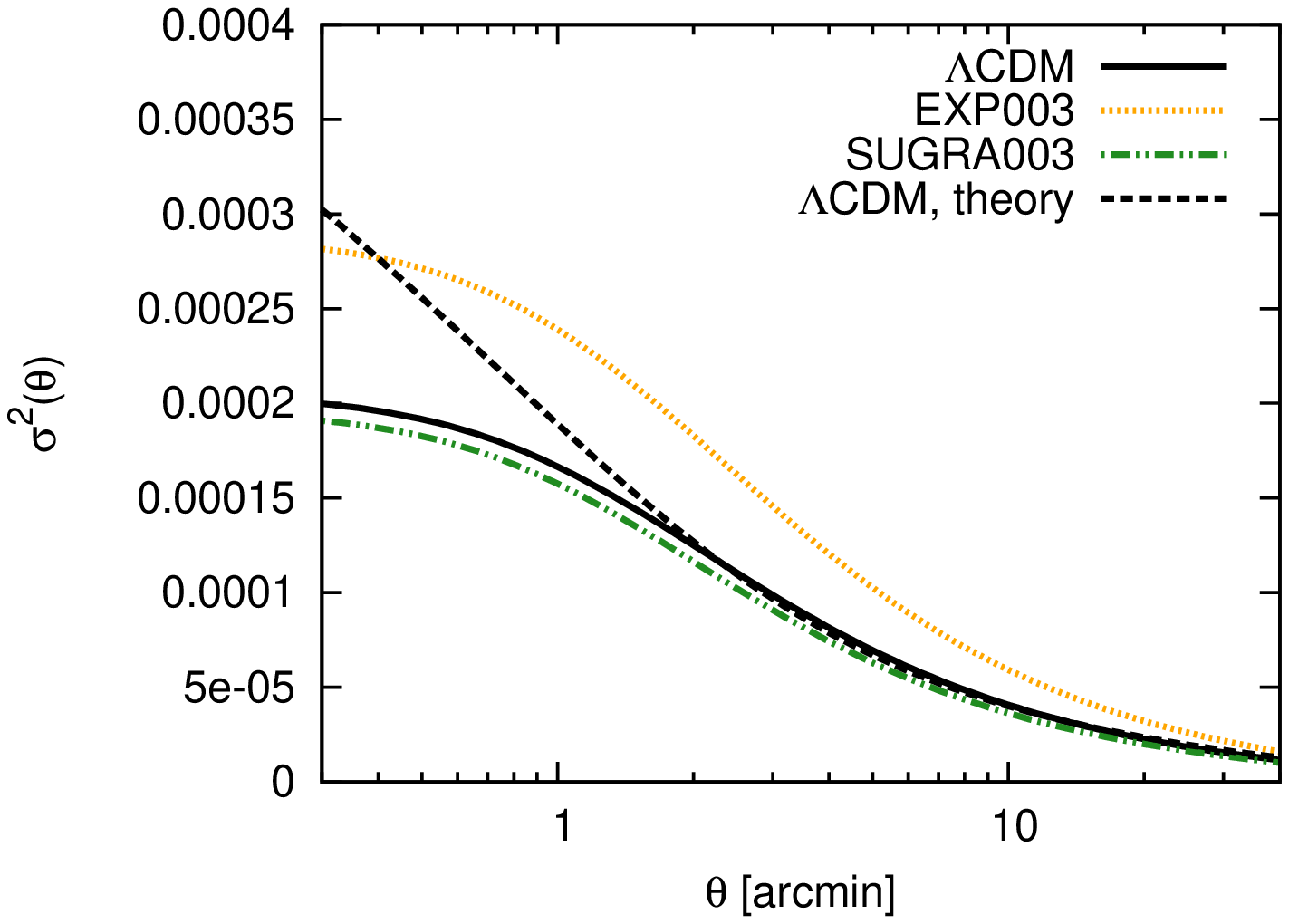}
 \includegraphics[width=0.3\textwidth,angle=-90]{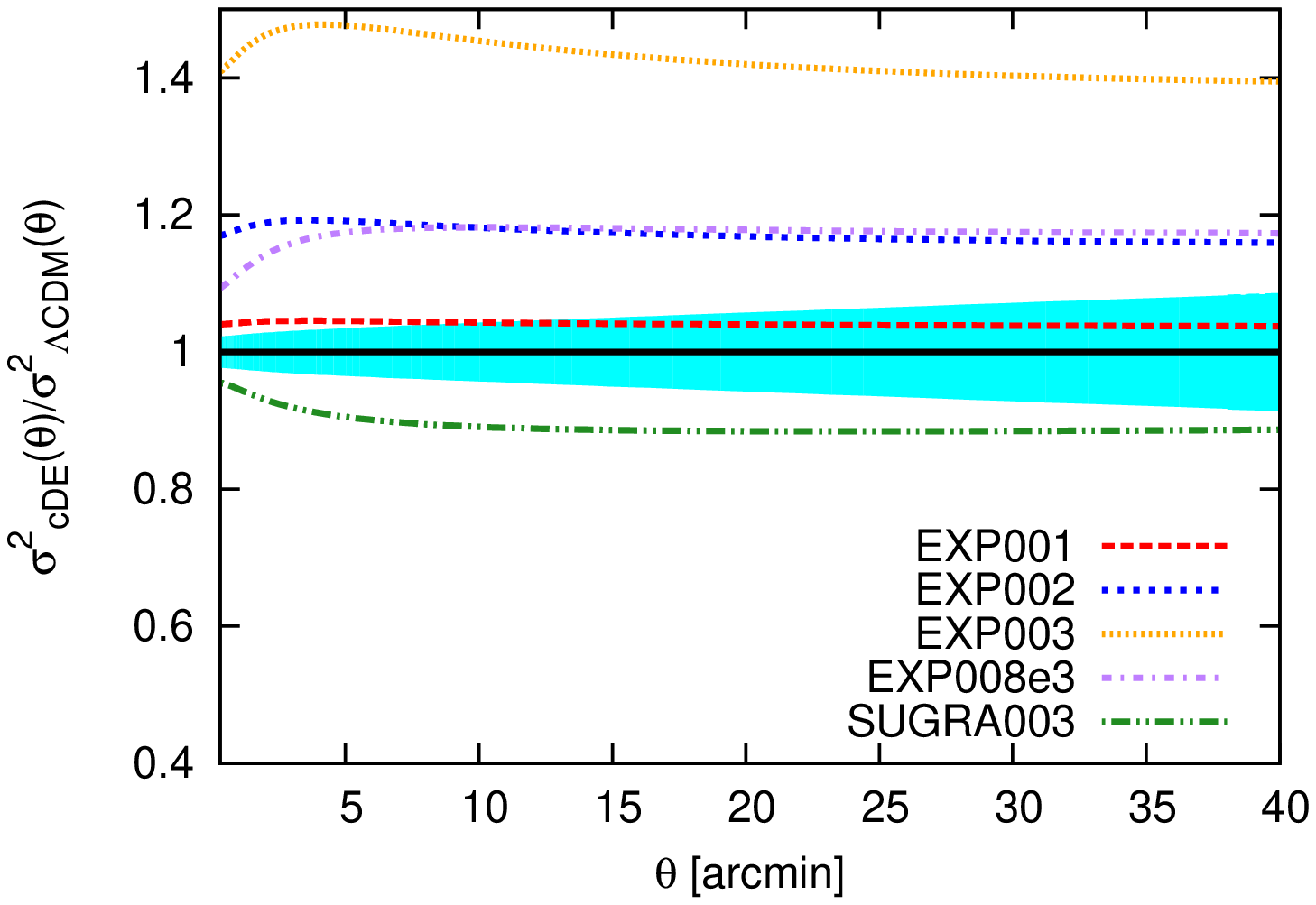}
 \caption{Shear in aperture. Upper panel: the results for the reference $\Lambda$CDM model and for the two most 
extreme coupled dark energy models (SUGRA003 and EXP003). For comparison we also show the analytical prediction for 
the $\Lambda$CDM model (dashed black line). Lower panel: the ratio between the coupled dark energy models and the 
$\Lambda$CDM model. Colour lines and styles are as in Fig.~\ref{fig:PSkappa}. The curves and the shaded region 
(shown only for the $\Lambda$CDM model) represent the average and the r.m.s. obtained from 100 different 
realizations, respectively.}
 \label{fig:gamma_var}
\end{figure}

In Fig.~\ref{fig:gamma_var} we show our results for the shear in aperture as a function of the angular scale 
$\theta$. In the upper panel we show a comparison between the values of the shear in aperture for the $\Lambda$CDM 
model (black solid line) and the two most extreme coupled dark energy models, the EXP003 model (orange dotted line) 
and the SUGRA003 model (green dashed-dot-dotted line). The power spectrum differences between the models translate to 
differences in the shear in aperture, but in an integrated, cumulative way.

For smaller apertures, the finite resolution of the simulations also becomes an issue, as can be seen by comparing 
with analytical predictions for the $\Lambda$CDM model (black dashed line). From Fig.~\ref{fig:gamma_var}, we see 
that simulations reliably reproduce the expected analytical result only for angles $\theta\geqslant 2$ arcmin, while 
on smaller scales the differences become substantial. For scales of the order of 0.3 arcmin, the lack of signal is 
about $\approx 30\%$. Our plot is very similar to that shown in \cite{Bartelmann2001}, their Fig.~19. The deficit is 
similar to what happens when the linear spectrum is used instead of the non-linear one. While our simulation is 
obviously fully non-linear, due to resolution effects, we lose some of the power on non-linear scales.

In the lower panel of Fig.~\ref{fig:gamma_var}, we show the ratio of the coupled dark energy models with respect to 
the reference model. Again, the shaded region represents the error bars obtained as r.m.s. of 100 different 
realizations. Error bars increase with increasing angular scale, since there are fewer independent patches in the map 
to average over. Since the simulation scatter is very small (shaded region), the EXP003 model could be easily 
distinguished from the $\Lambda$CDM model given such an observation. In general, the ratios have similar values to 
the ones found for the lensing power spectrum, and the ratio is approximately constant over the range of angular 
scales investigated in this work. Once again the different behaviour of the models is easily interpreted in terms of 
the different normalization of the matter power spectra (EXP models) and of the time evolution of the matter density 
parameter (the SUGRA003 model).

The reason why the ratio is approximately constant on all scales relates to the definition of the shear in aperture 
(Eq.~\ref{eqn:shear_aperture}). As noticed in Sect.~\ref{sect:PS}, the spectra are approximately a rescaled version 
of the $\Lambda$CDM model, therefore also its integral over the multipoles will be such that the variance is 
approximately a rescaled version of the $\Lambda$CDM expression. This is indeed confirmed in Fig.~\ref{fig:gamma_var}.

As in Section~\ref{sect:PS}, we wish to see whether the differences in the shear in aperture predictions are simply 
due to the higher normalization or we can observe some feature more directly reflecting the new physics. 
We again compared the EXP003 model with the predictions for a $\Lambda$CDM model having identical matter power 
spectrum normalization. Such a comparison shows a qualitative agreement on the ratios with respect to the reference 
$\Lambda$CDM model, including the peak in the ratio for $\theta\approx $ few arcmin. There is also a smaller 
impact from the feature seen in the power spectrum ratios, as we verified with a control ratio of the shear in 
aperture for two $\Lambda$CDM models with different matter power spectrum normalization. 
Nevertheless, the amplitude of the EXP003 model is lower than the amplitude of the $\Lambda$CDM model with analogous 
normalization of the matter power spectrum. 
This is in agreement with our finding for the convergence power spectrum (see Fig.~\ref{fig:LCDMps8}) and it is due 
to the friction term in the equations of motion. We refer to \cite{Baldi2012} for a further discussion of the subject.

\subsection{Shear correlation Function}\label{sect:CF}
Another counterpart of the lensing power spectrum discussed in Sect.~\ref{sect:PS} is the shear correlation 
function defined as
\begin{equation}
 \xi_{+}(\theta)=\int_0^{\infty}d\ell\frac{\ell}{2\pi}P_{\kappa}(\ell)J_{0}(\ell\theta)\;,
\end{equation}
where $J_{0}(x)$ is the Bessel function of order zero, $P_{\kappa}(\ell)$ the effective convergence power spectrum 
and $\theta$ the angular distance between the correlated sources.  Note that the kernel in the integrand is different 
from the shear in aperture statistics, therefore these can be compared only qualitatively.

A detailed study of the shear correlation function was performed by \cite{Beynon2012}, which we refer to for more 
details. However there is a substantial difference compared to that work: our simulations assume that all the 
sources are at $z_{\rm s}=1$, while in \cite{Beynon2012} sources follow a different redshift distribution according 
to the different weak lensing survey the prediction is made for. Moreover their shear correlation function is 
presented only for the models EXP001, EXP002 and EXP003. We can therefore only make a qualitative comparison between 
the two different analyses.

\begin{figure}
 \centering
 \includegraphics[width=0.3\textwidth,angle=-90]{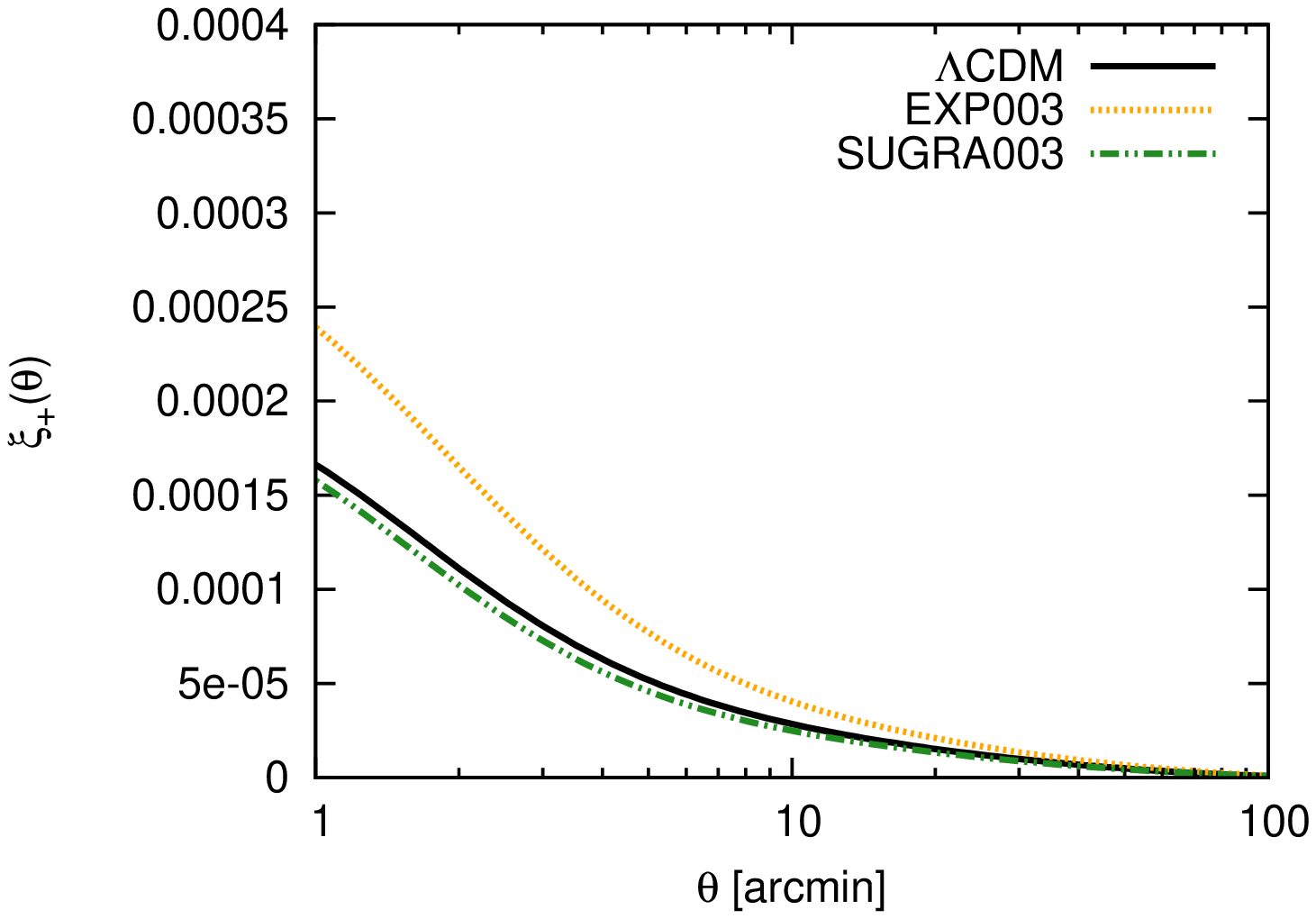}
 \includegraphics[width=0.3\textwidth,angle=-90]{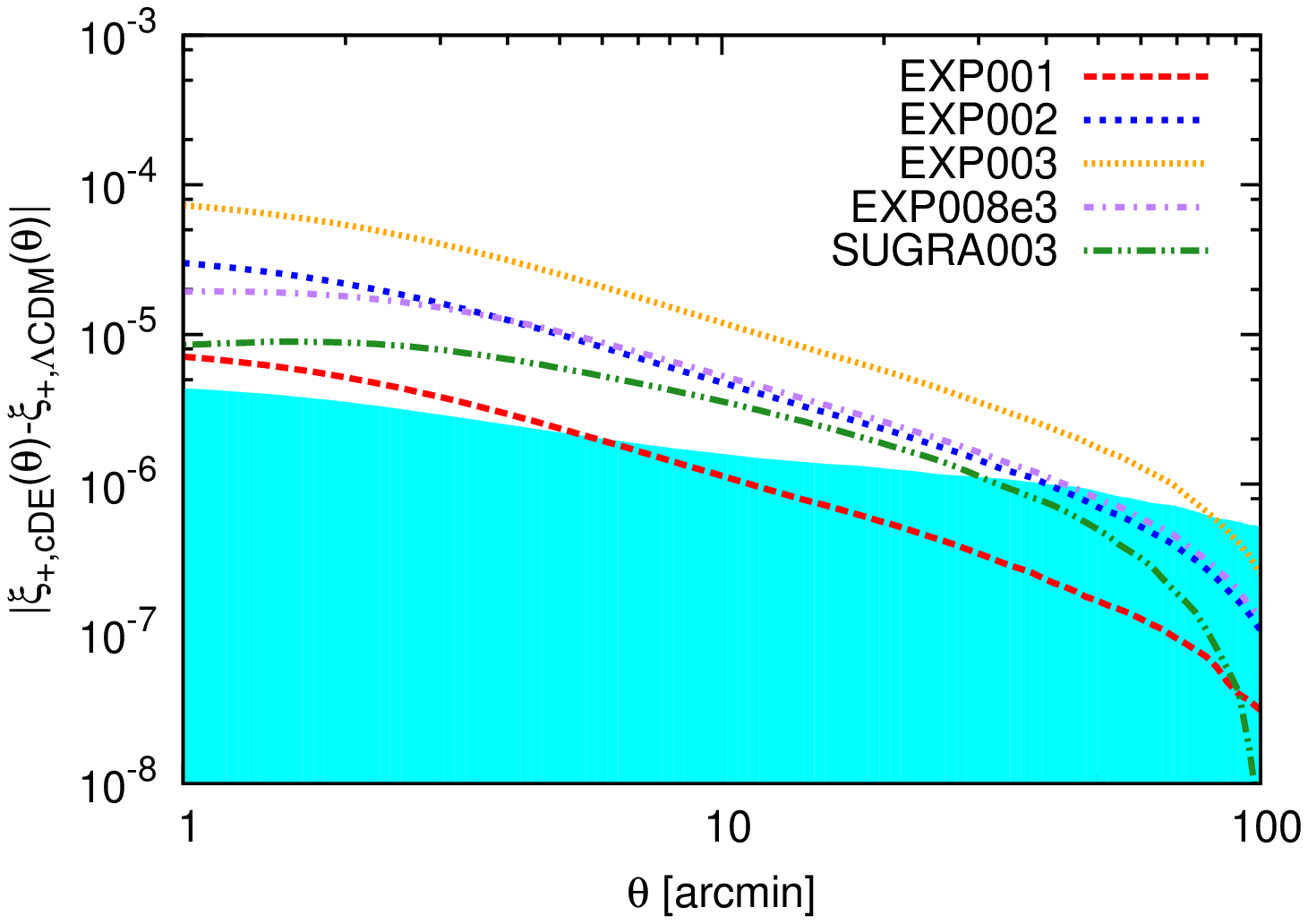}
 \caption{Shear correlation function. Upper panel: the results for the reference $\Lambda$CDM model and for the two 
most extreme coupled dark energy models (SUGRA003 and EXP003). Lower panel: absolute value of the difference between 
the coupled dark energy models and the $\Lambda$CDM model. Colour lines and styles are as in Fig.~\ref{fig:PSkappa}. 
The curves and the shaded region (shown only for the $\Lambda$CDM model) represent the average and the r.m.s. 
obtained from 100 different realizations, respectively.}
 \label{fig:gamma_cf}
\end{figure}

We present the correlation functions in Fig.~\ref{fig:gamma_cf} as a function of the angular scale $\theta$. In the 
upper panel we 
show a comparison between the values of the shear correlation function for the $\Lambda$CDM model (black solid line) 
and the two most extreme coupled dark energy models, the EXP003 model (orange dotted line) and the SUGRA003 model 
(green dashed-dot-dotted line).   
As expected, with respect to the reference $\Lambda$CDM model, we see an excess (a lack) of correlation for the 
EXP003 (SUGRA003) model. Once again, we can explain this result in terms of the different matter density evolution 
(SUGRA003 model) and of the different matter power spectrum normalisation (EXP003 model).

In the lower panel of Fig.~\ref{fig:gamma_cf}, we present the differences of the shear correlation function 
$\xi_{+}(\theta)$ between the coupled dark energy models and the $\Lambda$CDM model. The shaded region represents the 
1-sigma error bar as obtained averaging over 100 realizations. In agreement with \cite{Beynon2012}, we see that 
errors decrease with increasing the correlation angle. This is expected since there are more objects to average over. 
The amplitude of the r.m.s. errors is different from \cite{Beynon2012}, since ours is based on the different 
realizations performed, while the value presented in \cite{Beynon2012} refers to the discriminatory power of the 
specific survey.

With respect to \cite{Beynon2012}, our predictions for the shear correlation function are somewhat higher. This is 
expected since in our simulations all the sources are at the same redshift. The differences in the shear correlation 
functions arising from different  
redshift distributions of the sources is not a simple constant, but it is a function of the angular scale. 
In addition, the behaviour at small angular scales is due to resolution effects that lead to a loss of power.

To summarise, our results are in good qualitative agreement with \cite{Beynon2012}. Models with higher power spectrum 
normalisation show a higher amplitude of the shear correlation function while the SUGRA model presents a deficit in 
the signal (since we take the absolute values, the SUGRA model lies above the EXP001 model). The trend closely 
follows what found for the study of the power spectrum and of the shear in aperture. The model EXP001 is once again 
barely above the 1-sigma error bars, making it therefore difficult to detect (differently from what found in 
\cite{Beynon2012}), but on a wide range of angular scales the EXP003 will be clearly identified. Models EXP002 and 
EXP008e3 behave in a very similar way, analogously to what found for the power spectrum and shear in aperture. All 
the other models are within the error bars for $\theta\gtrsim 30-40$ arcmin, once again differently from 
\cite{Beynon2012}. Taking into account that as shown in Fig.~5 of \cite{Beynon2012}, using Halofit \citep{Smith2003} 
introduces errors of the same order of magnitude of the intrinsic differences between the models, we can conclude that 
raytracing simulations are an important tool in studying this class of models.

\begin{figure*}
 \includegraphics[width=0.3\textwidth,angle=-90]{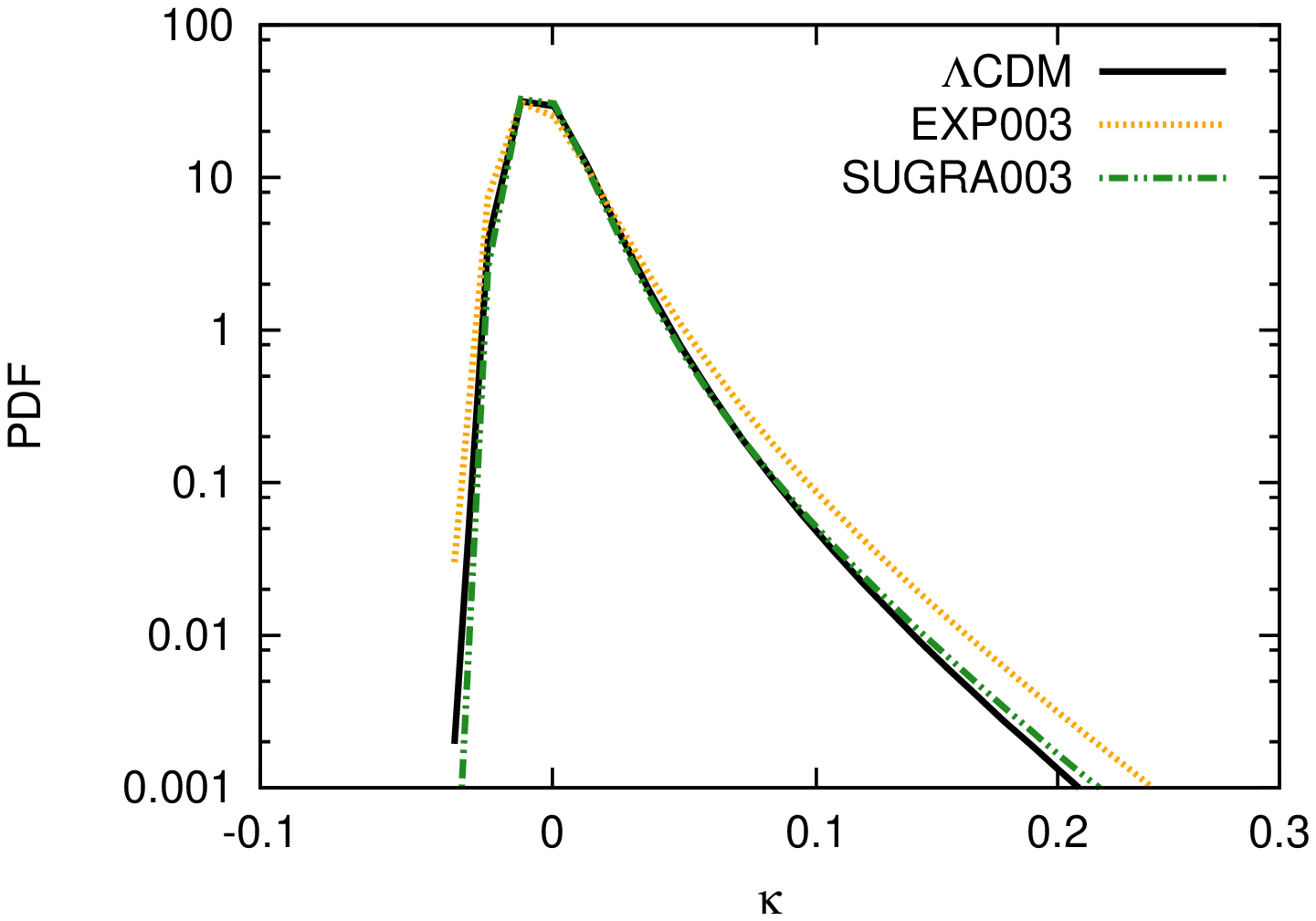}
 \includegraphics[width=0.3\textwidth,angle=-90]{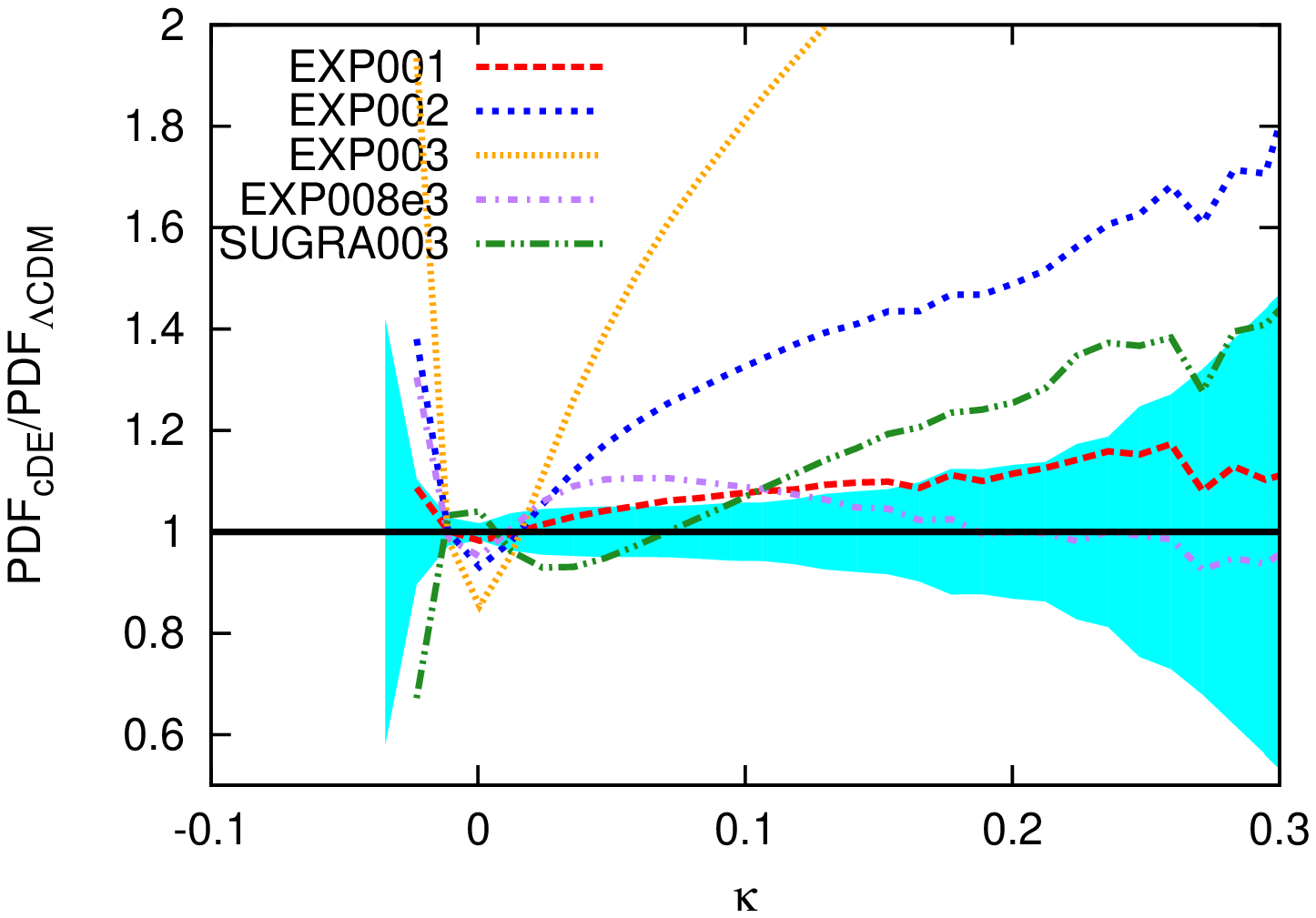}
 \includegraphics[width=0.3\textwidth,angle=-90]{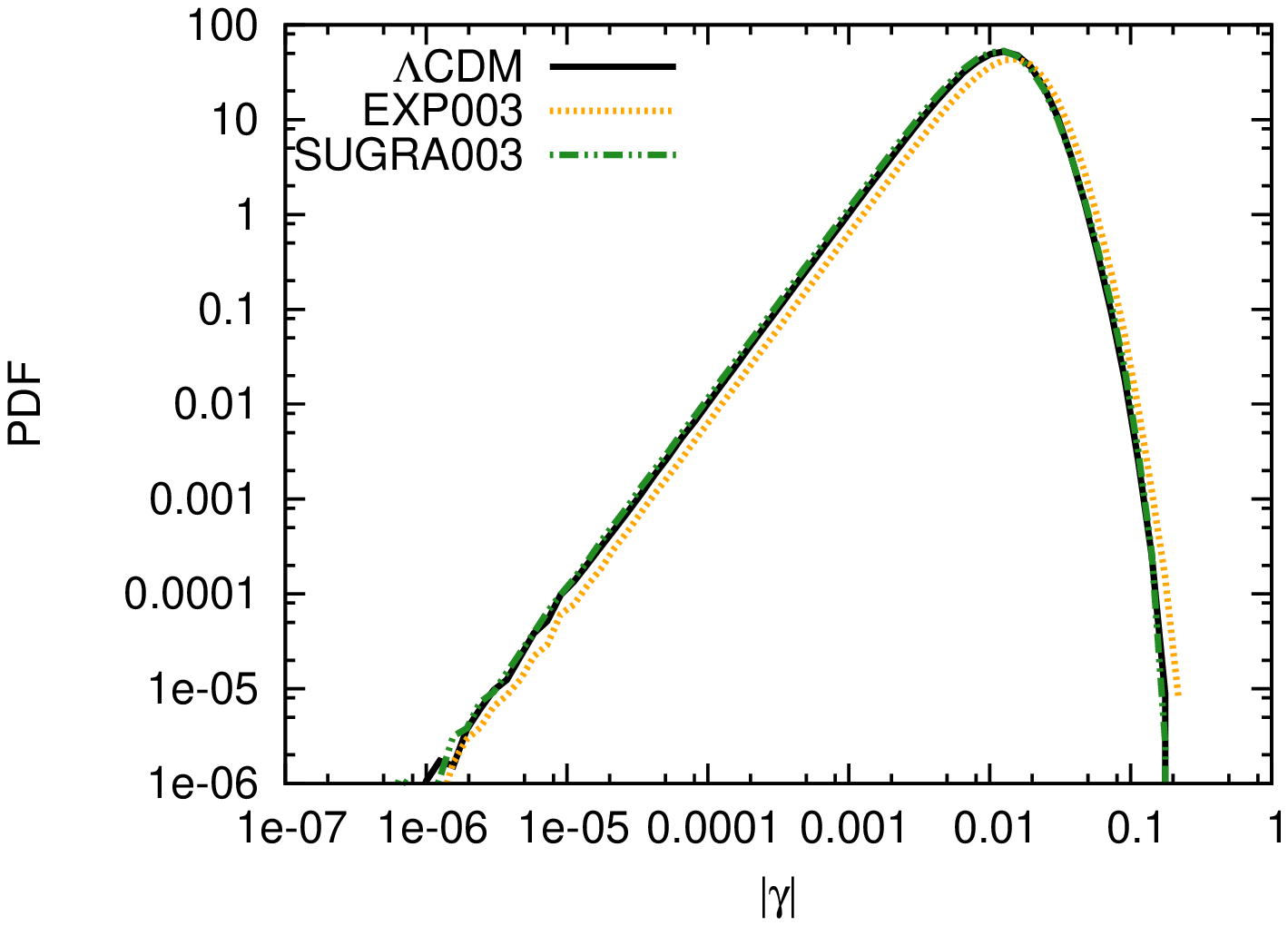}
 \includegraphics[width=0.3\textwidth,angle=-90]{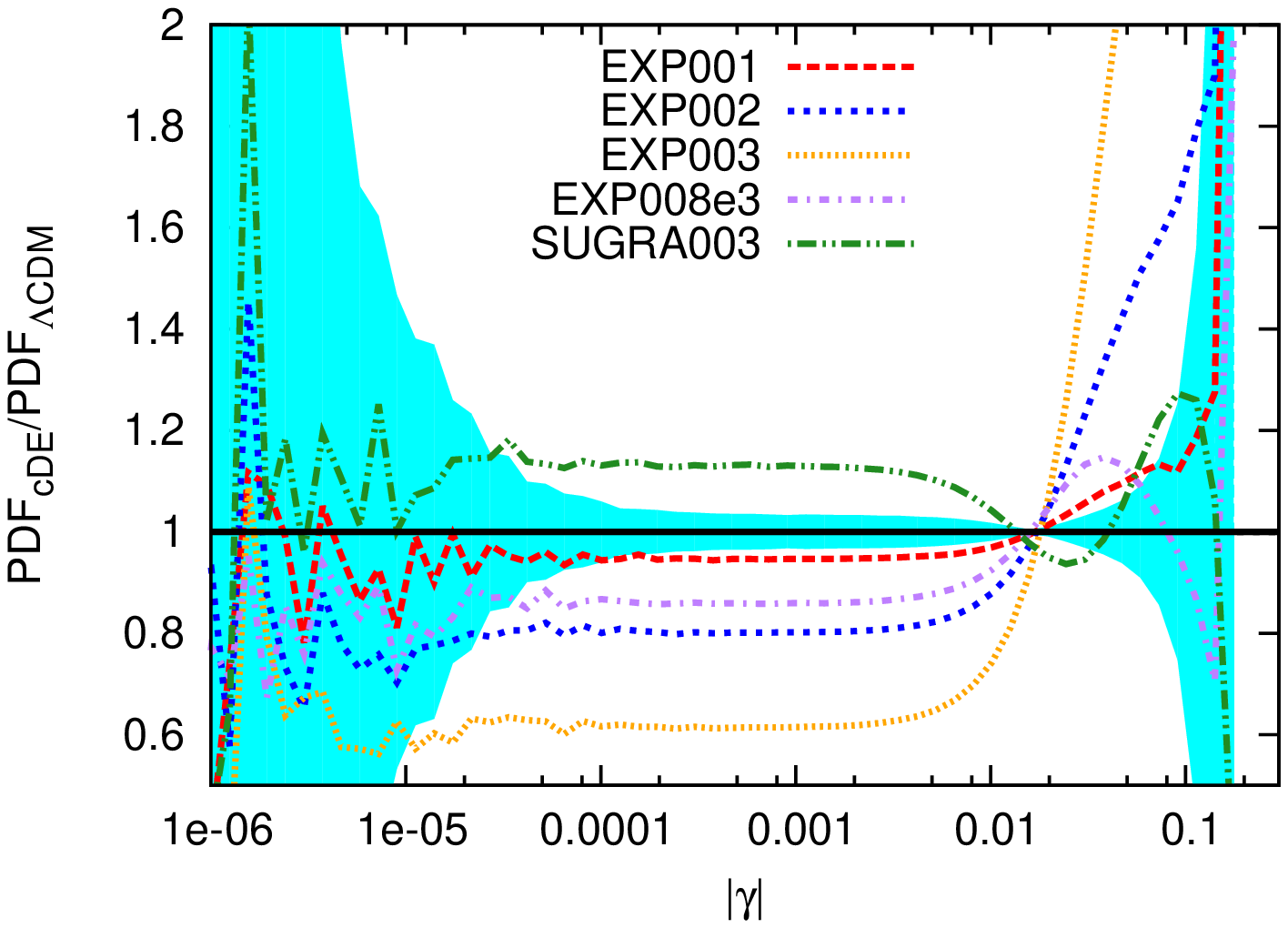}
 \includegraphics[width=0.3\textwidth,angle=-90]{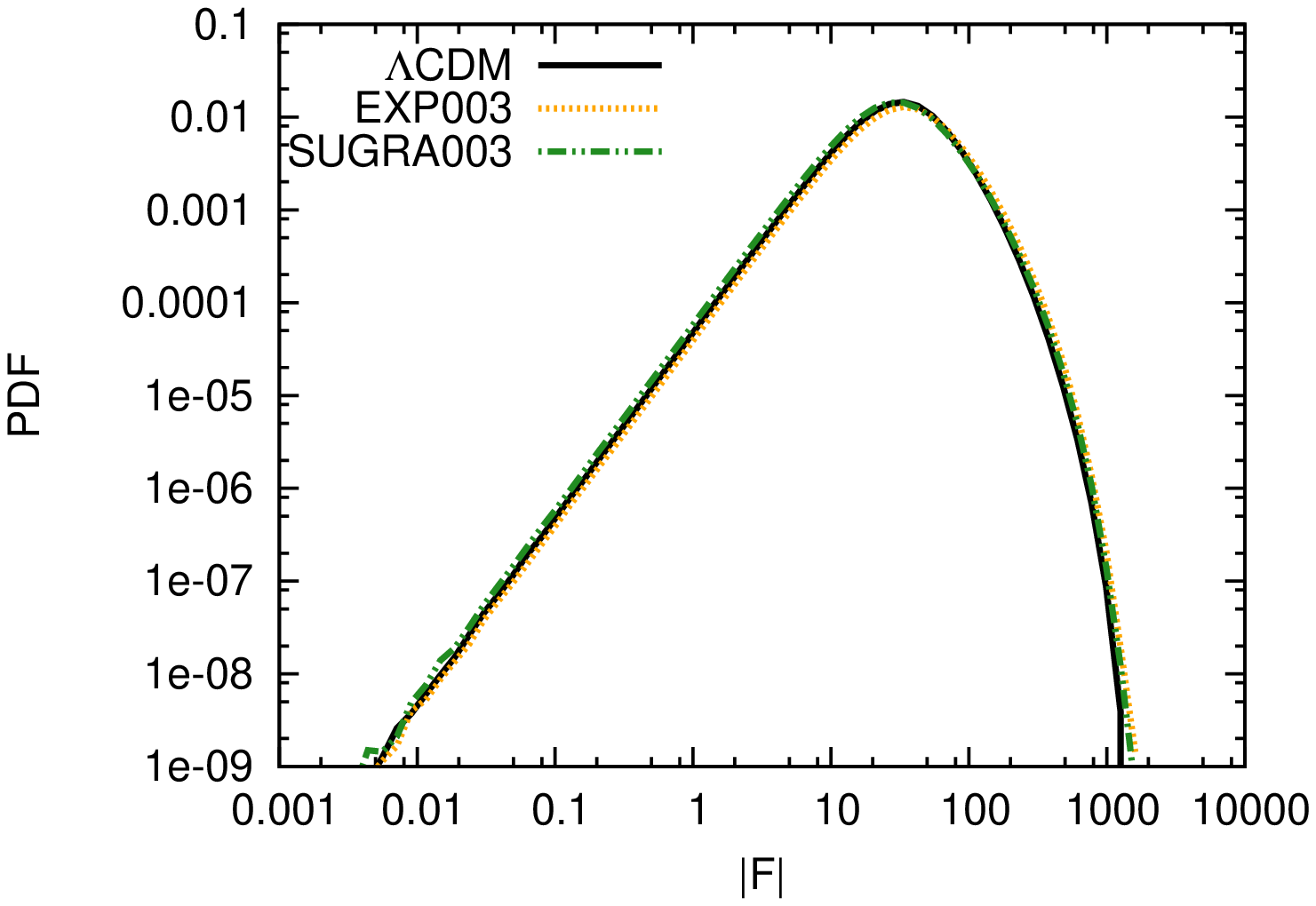}
 \includegraphics[width=0.3\textwidth,angle=-90]{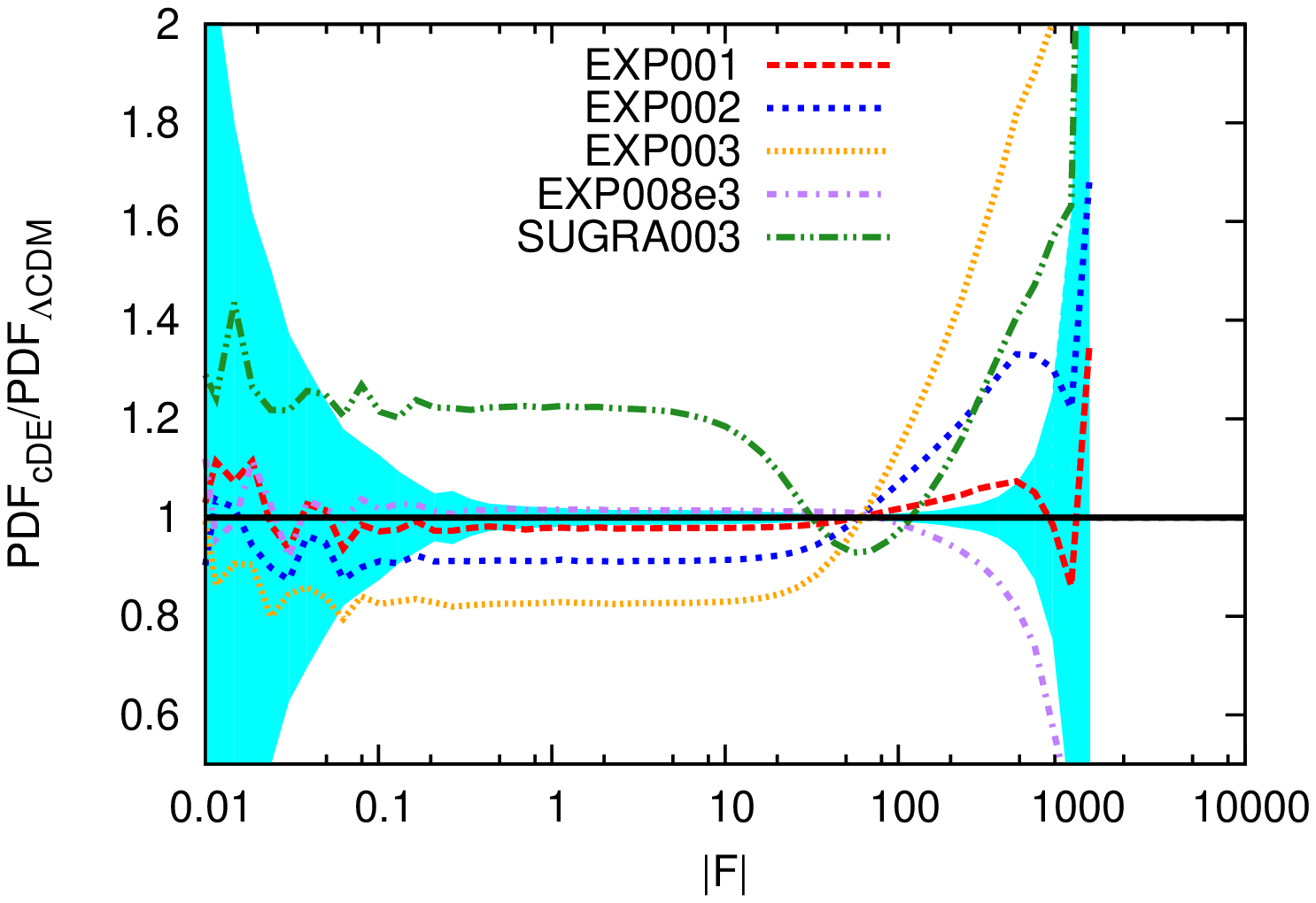}
 \includegraphics[width=0.3\textwidth,angle=-90]{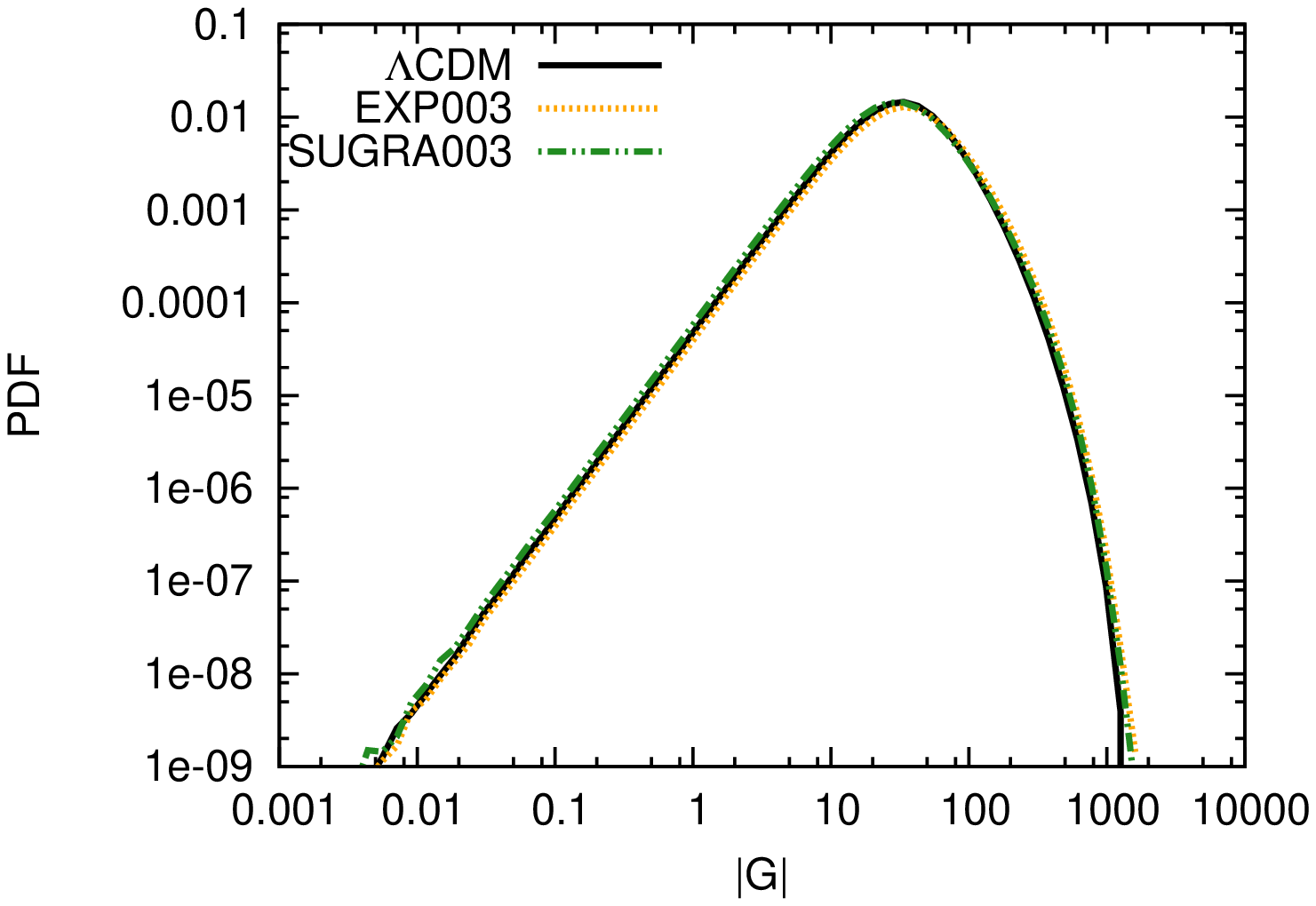}
 \includegraphics[width=0.3\textwidth,angle=-90]{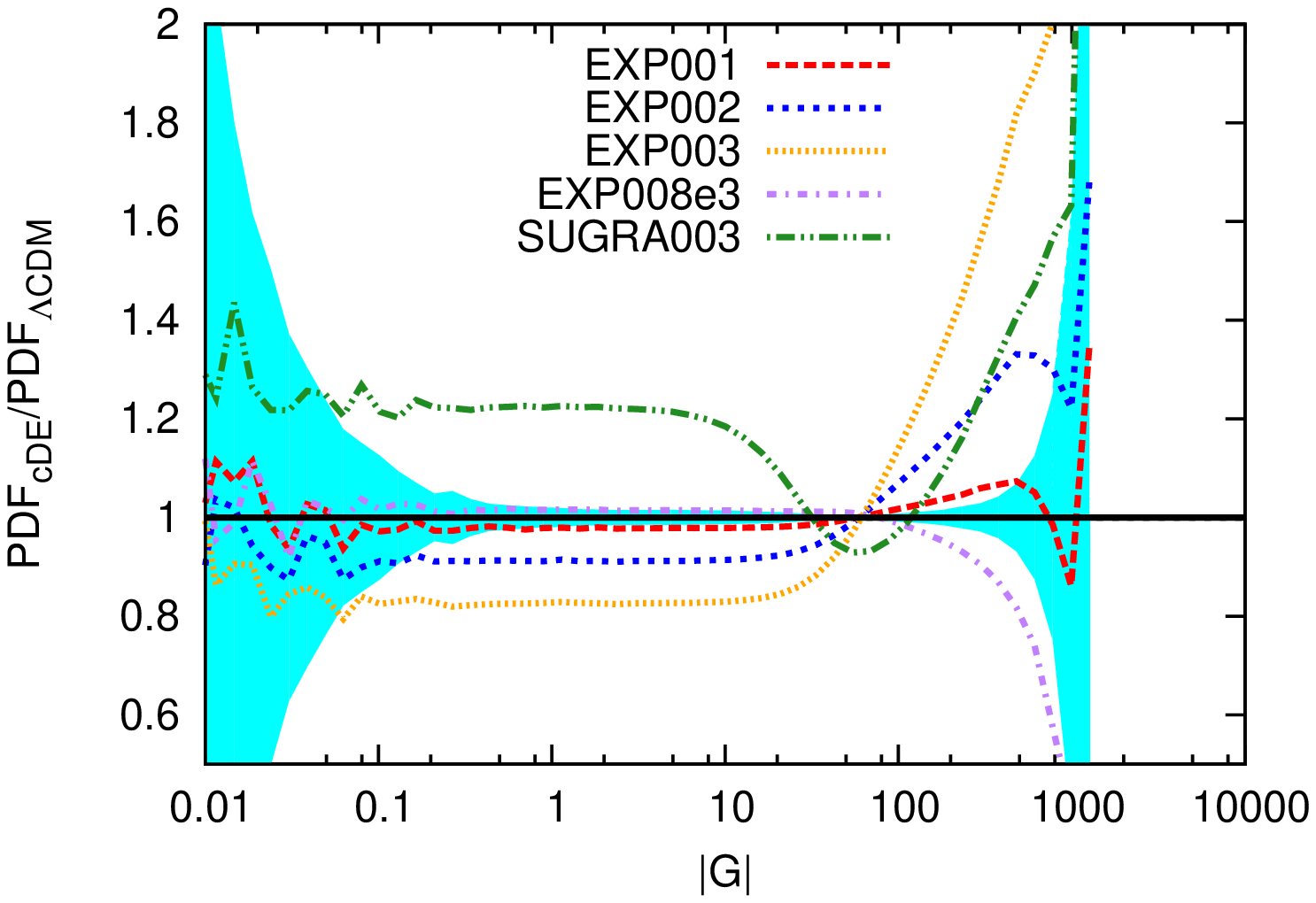}
 \caption{PDF for several lensing quantities analysed in this work. From top to bottom: effective convergence 
$\kappa$, modulus of the shear $\gamma$, 1- and 3-flexion ($F$ and $G$, respectively). Left panels show the results 
for the reference $\Lambda$CDM model and for the two most extreme coupled dark energy models (SUGRA003 and EXP003). 
Right panels show the ratio between the coupled dark energy models and the $\Lambda$CDM model. Colour lines and 
styles are as in Fig.~\ref{fig:PSkappa}. The curves and the shaded region (shown only for $\Lambda$CDM model) 
represent the average and the r.m.s. obtained from 100 different realizations, respectively.}
 \label{fig:PDF1}
\end{figure*}

\subsection{Probability Distribution Function}\label{sect:PDF}
While the power spectrum and shear in aperture fundamentally reflect the same statistical information, it is 
interesting to explore whether non-Gaussianity of the lensing statistics can help distinguish between the different 
physics. We explore this first by examining the full one-point probability distribution function (PDF), and discuss 
moments of the PDF in the next section. We limit our discussion to quantities that can be effectively observed, in 
particular to the effective convergence, the (modulus of the) shear, 1- and 3-flexion ($F$ and $G$) and finally the 
magnification.

To infer the PDF from our simulated lensing maps, we first establish the absolute minimum and maximum of the 
maps for a given quantity, then we bin the values of the maps in this interval. Since each map has a different range 
of values, binning the pixels in a range enclosed by the absolute minima and maxima allows us to compute the ratio 
between the different models straightforwardly, without the need to interpolate or extrapolate the numerical PDF.

As it is apparent from Fig.~\ref{fig:PDF1}, for the modulus of the shear and of the two flexions, differences between 
the coupled dark energy models and the $\Lambda$CDM model are of the order of 20\%-40\%, and percentage differences 
for the two flexions are similar to those for shear. 
In particular the model EXP003 now shows differences of the order of 40\% with respect to the $\Lambda$CDM model. 
We notice that error bars are relatively small for all the quantities, but at the two extremes representing 
relatively rare extreme underdense and overdense regions. 
As seen above, the models most significantly different from the reference one are the SUGRA003 and the EXP003, due 
to the lower effective matter density parameter for the first and the higher matter power spectrum normalization for 
the second. 
The shear PDF instead is more sensitive to the matter power spectrum normalization; we see that differences can be up 
to 40\% and approximately constant over a few decades of the shear values.

Regarding the effective convergence, we see in the top panel of Fig.~\ref{fig:PDF1} that differences between the 
models grow largest in the high convergence tail. Note that unlike the shear and flexion moduli, the convergence can 
take both positive and negative values and it can be well fitted by a lognormal distribution with mean 
$\kappa_0=0.04$ and variance $\sigma=0.35$ [following the notation of \cite{Taruya2002} and \cite{Hilbert2011}]. 
For models with an increasing power spectrum normalization, differences become more pronounced, particularly for the 
most extreme EXP003 model, which is characterised by a very high $\sigma_8$. 
The models SUGRA003 and EXP008e3 are quite interesting; having a slightly lower normalization than the $\Lambda$CDM 
model, the SUGRA003 model has an excess of high convergence points. 
On the other hand, the EXP008e3 model, despite having a significantly higher normalisation, shows no change in the 
tail with respect to the fiducial model. These demonstrate that the friction terms can have a very pronounced effect 
on the formation of non-linear structures. This was also seen in small scale power spectrum in Fig.~\ref{fig:PSkappa}.

While gravitational lensing preserves the surface brightness, this is not the case for the apparent solid angle of a 
source. The magnification $\mu$, defined as the ratio of the image area to the source area, can be expressed in terms 
of the shear $\gamma$ and effective convergence $\kappa$ via the relation
\begin{equation}\label{eqn:mu}
 \mu=\frac{1}{(1-\kappa)^2-\gamma^2}\;.
\end{equation}
Recently the magnification has become an active research field for cosmology due to its power in complementing shear 
studies 
\citep[][]{Bernstein2002,vanWaerbeke2010a,vanWaerbeke2010b,Hildebrandt2011,Ford2012,Casaponsa2013,
Heavens2013,Hildebrandt2013}.

\begin{figure}
 \centering
 \includegraphics[width=0.3\textwidth,angle=-90]{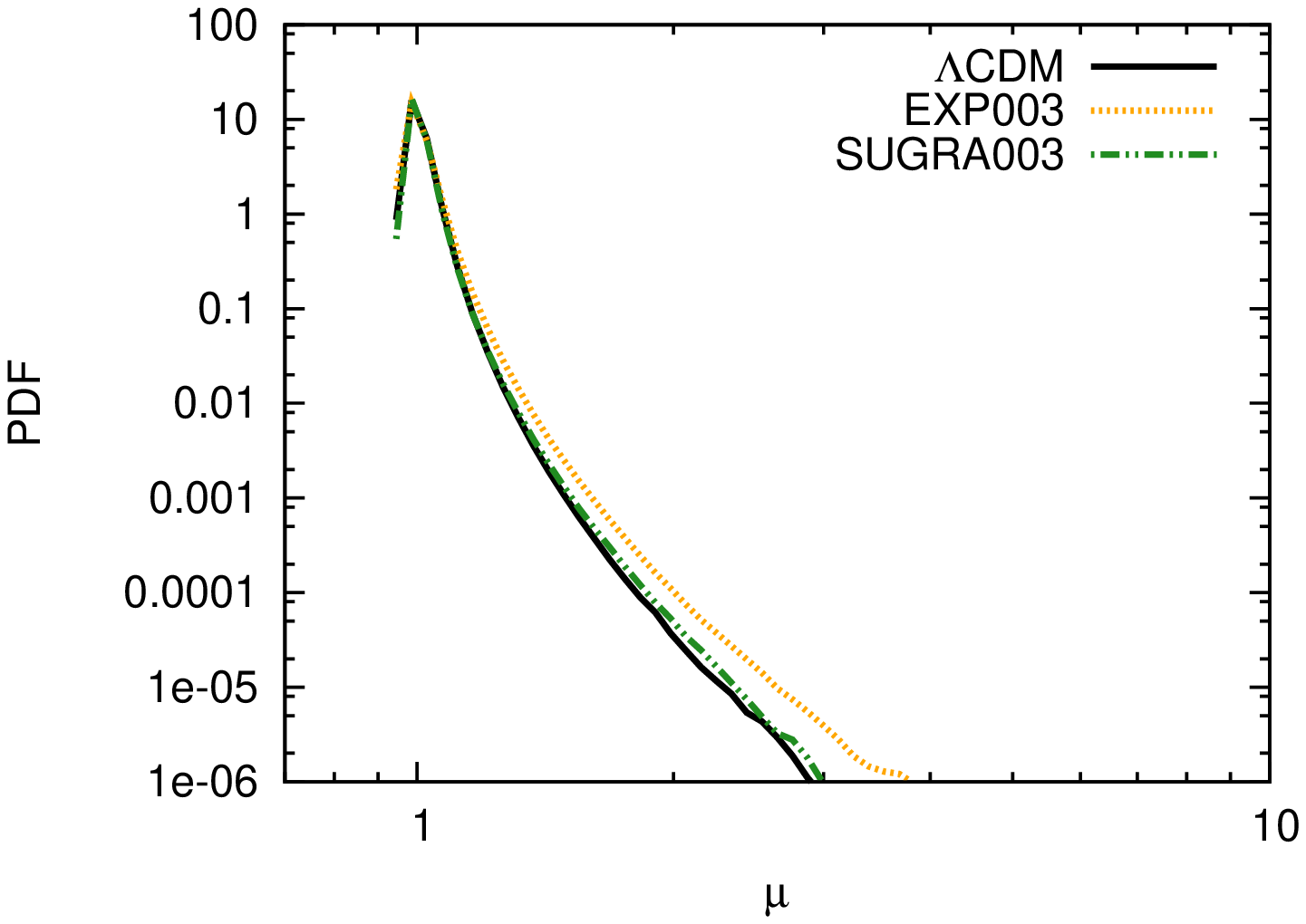}
 \includegraphics[width=0.3\textwidth,angle=-90]{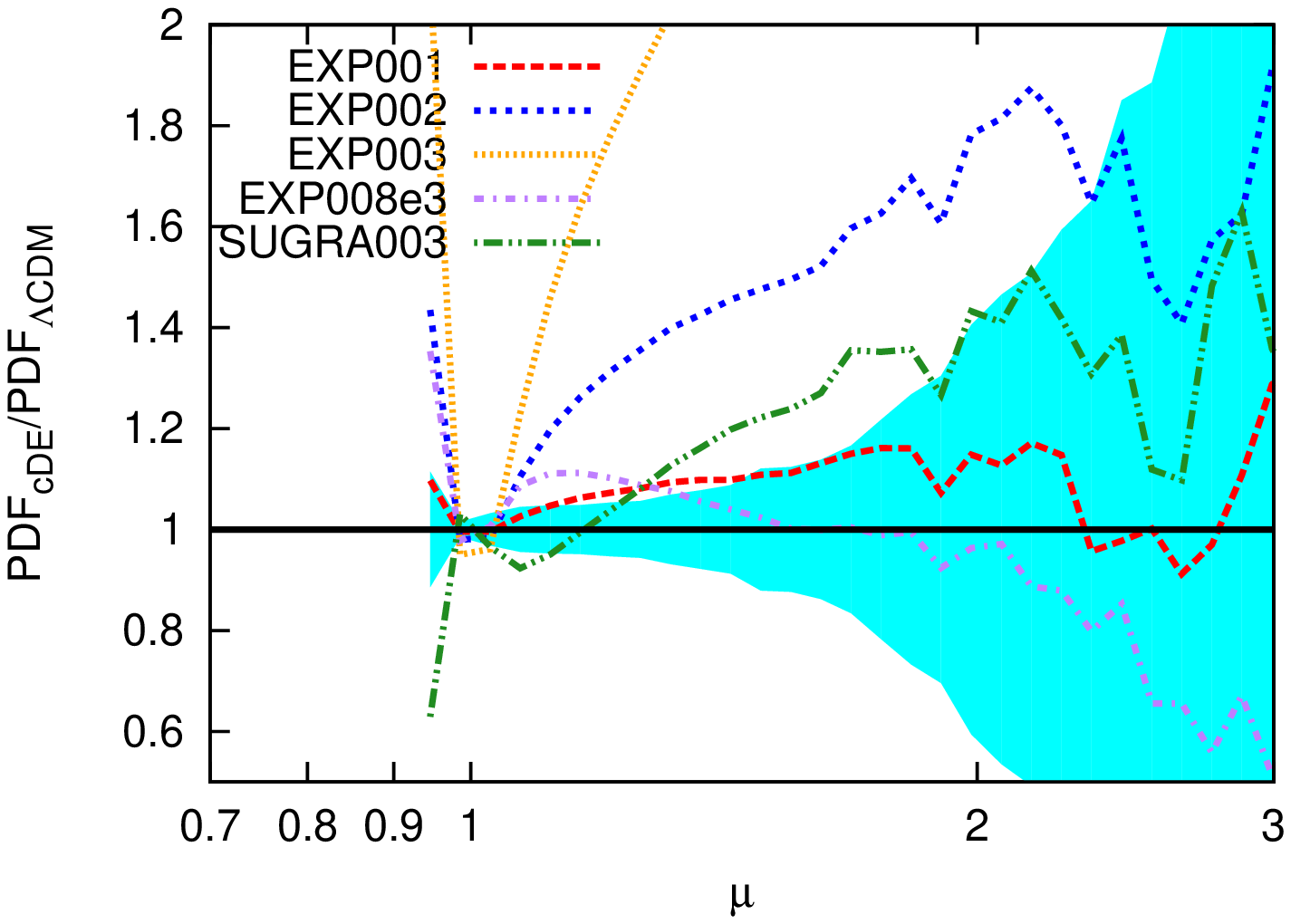}
 \caption{PDF for the cosmic magnification. The upper panel shows the results for the reference $\Lambda$CDM model 
and for the two most extreme coupled dark energy models (SUGRA003 and EXP003). The lower panel shows the ratio 
between the coupled dark energy models and the $\Lambda$CDM model. Colour lines and styles are as in 
Fig.~\ref{fig:PSkappa}. The curves and the shaded region (shown only for $\Lambda$CDM model) represent the average 
and the r.m.s. obtained from 100 different realizations, respectively.}
 \label{fig:PDFmu}
\end{figure}

Examining Fig.~\ref{fig:PDFmu}, we notice that magnification can also be an excellent discriminant between different 
models, even if in this case error bars are much bigger than before at the high magnification tail. In particular, as 
noticed with the effective convergence case, the model EXP003, having a much higher $\sigma_8$, makes a higher 
magnification of the source more probable. This is easily understood by considering the relation between the 
magnification and the effective convergence, $\mu\simeq 1+2\kappa$ (valid at first order) when both the shear and the 
effective convergence are small. Therefore higher values of the convergence also imply higher values for the 
magnification. At second order, taking into account both the effective convergence and the shear, the relation between 
the magnification and these two quantities becomes \citep{Menard2003,Takahashi2011,Marra2013}
\begin{equation}\label{eqn:mukg}
\mu \simeq 1+2\kappa+3\kappa^2+\gamma^2+\mathcal{O}(\kappa^3,\gamma^3)\;.
\end{equation}
The magnification $\mu$, up to second order, depends on the convergence $\kappa$ and its square ($\kappa^2$) and on 
the square of the modulus of the shear ($\gamma^2$). In Fig.~\ref{fig:PDF1} we saw that both the effective convergence 
and the shear are sensitive to the background cosmological model; therefore we cannot neglect the contribution coming 
from the shear. While at small shear and convergence we can relate the two PDFs via the expression 
\citep{Takahashi2011}
\begin{equation}
 \frac{dP_{\mu}}{d\mu}=\frac{(1-\kappa)^3}{2}\frac{dP_{\kappa}}{d\kappa}\;,
\end{equation}
where $dP_{\mu}/d\mu$ and $dP_{\kappa}/d\kappa$ are the PDF's of the magnification and of the effective convergence, 
respectively, this is no longer accurate at larger values of convergence and shear. This is reflected in comparing 
the top panels of Fig.~\ref{fig:PDF1} with Fig.~\ref{fig:PDFmu}.

As for the power spectrum, it is interesting to understand which PDF differences are simply due to the different 
normalisation of the models and which are more intrinsic. 
To evaluate the PDF of the effective convergence or of the shear, two different approaches have been 
followed in the literature. On one hand, perturbation theory techniques 
\citep[see e.g.][]{Munshi2000,Taruya2002,Valageas2000a,Valageas2000b,Menard2003,Valageas2004a,Valageas2004b,
Takahashi2011} and the halo model \citep{Takada2003c} were exploited to analytically infer the PDF of the effective 
convergence $\kappa$, the modulus of $\gamma$ and of the magnification $\mu$; on the other hand, with the help of 
N-body simulations, numerical fits to the PDF of the effective convergence were determined, so as to have a quick 
recipe when cosmological parameters have to be changed, for example the matter density $\Omega_{\rm m,0}$ and the 
matter power spectrum normalization $\sigma_8$ \citep[see e.g.][]{Hilbert2011,Marra2013}.

\begin{figure}
 \centering
 \includegraphics[width=0.3\textwidth,angle=-90]{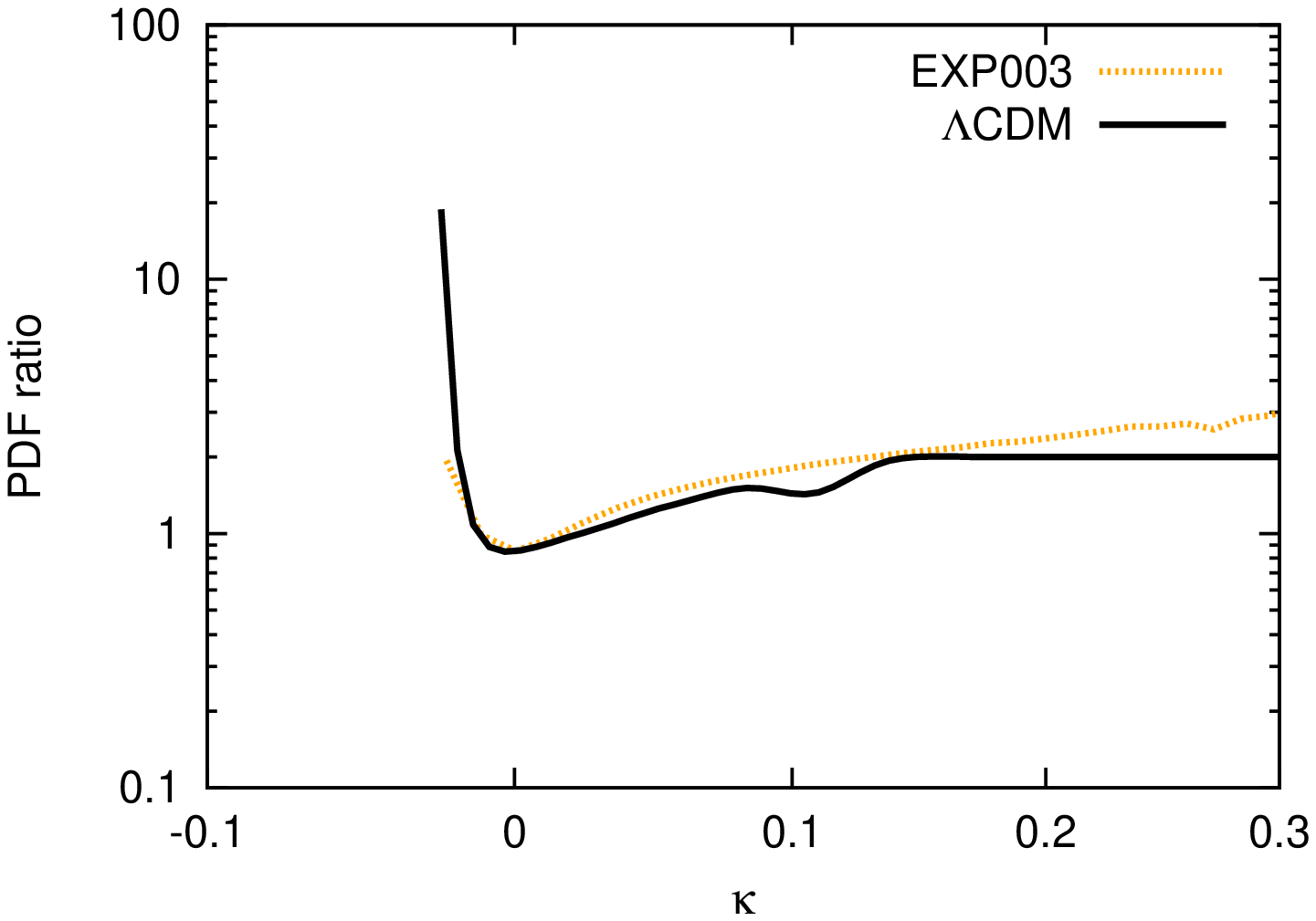}
 \includegraphics[width=0.3\textwidth,angle=-90]{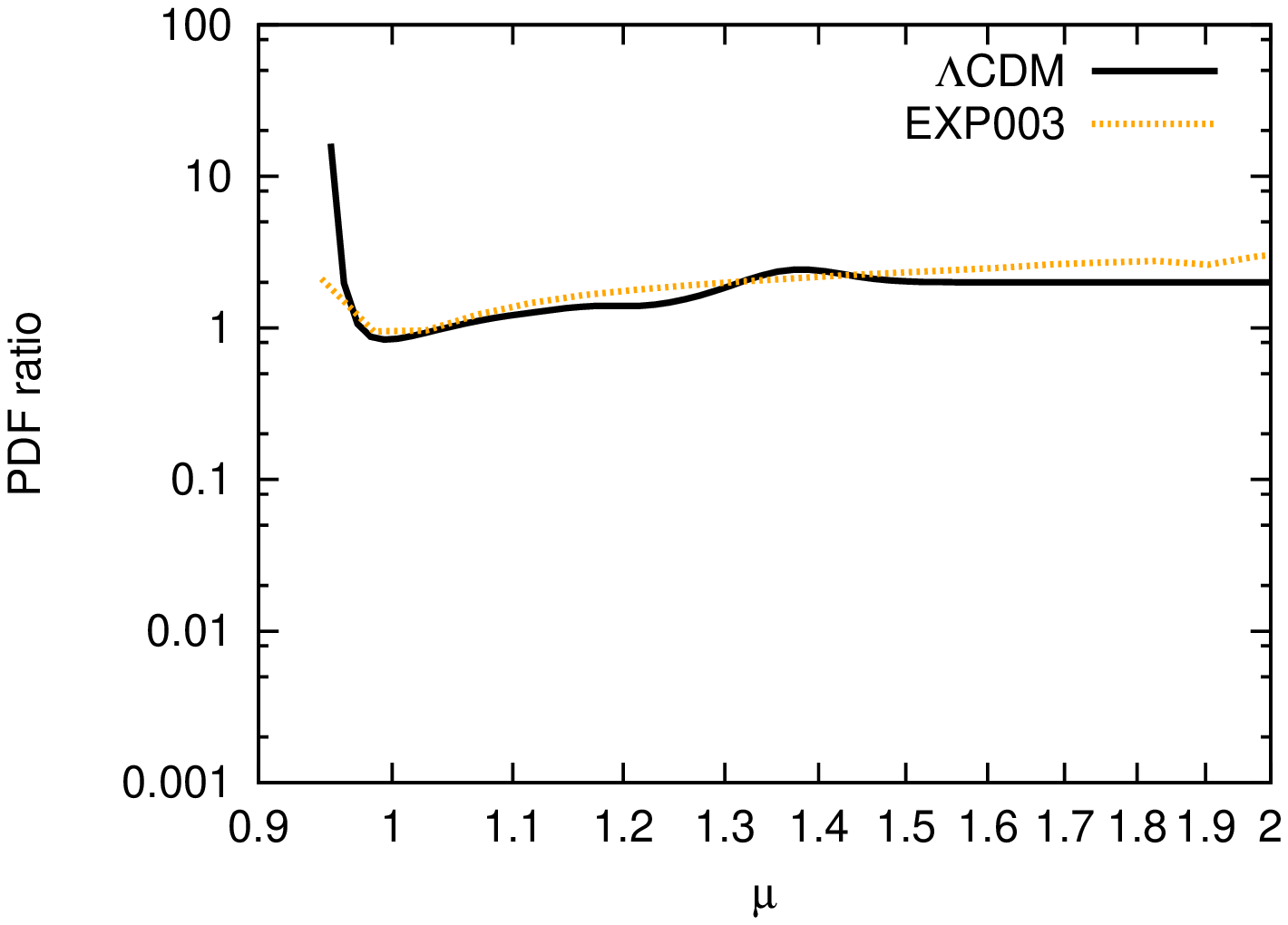}
 \caption{Ratio between the PDF of the EXP003 (orange dotted line) and the reference $\Lambda$CDM model. For 
comparison we show the same ratio obtained with the turboGL code for a $\Lambda$CDM model having the same 
normalization $\sigma_8=0.967$ as the EXP003 model (black solid line). The upper and lower panels refer to the 
effective convergence $\kappa$ and to the magnification $\mu$, respectively.}
 \label{fig:turboGL}
\end{figure}

For this work we use the output of the turboGL code\footnote{http://www.turbogl.org/} 
\citep{Kainulainen2009,Kainulainen2011}. The turboGL code is based on the stochastic approach to cumulative weak 
lensing and on generating stochastic configurations of halos along the line of sight, or along the 
photon geodesic from the source to the observer. Halos that model virialised structures are described by a 
Navarro-Frenk-White density profile \citep[][]{Navarro1996,Navarro1997}, filaments as non-uniform cylindrical objects. 
In addition, the modelling takes into account the fact that most of the cosmic volume is occupied by voids while most 
of the mass is in virialised structures and filaments. 
We show the comparison for the PDF of the effective convergence and of the magnification in Fig.~\ref{fig:turboGL} 
with different normalization of the matter power spectrum for the $\Lambda$CDM and EXP003 model.

It is apparent that the differences between the EXP003 and the $\Lambda$CDM model can be entirely explained in terms 
of the different normalization of the matter power spectrum. The range we can use is however limited, due to the 
fact that the raytracing procedure underestimates the true PDF for the effective convergence and magnification: this 
is due to the limited pixel resolution and mass assignment to create the lens planes \citep{Killedar2012}. In 
addition, the tail of the two distributions are not very well sampled, therefore we can not draw any conclusion on 
the fact that for high values of the effective convergence (and hence magnification), the two curves show small 
differences.

\subsection{Mean, Median, Variance, Skewness \& Kurtosis}\label{sect:high_order}
Signatures of coupling between dark energy and dark matter can be more easily quantified by considering higher order 
moments of the probability distribution function. The probability distribution function (Sect.~\ref{sect:PDF}) 
represents the one-point distribution, while the power spectrum (Sect.~\ref{sect:PS}) and the shear in aperture 
(Sect.~\ref{sect:shear_aperture}) represent second-order moments.
 
Next we will focus on the mean, the median, the variance, the skewness and the kurtosis of the PDF of the effective 
convergence, and by considering these at varying resolutions we effectively include the effect of spatial 
correlations.
 
Unfortunately, these statistical quantities are often affected by large errors which make their use on real data more 
difficult.

The mean $\mu_1$, the variance $\mu_2$, the skewness $\mu_3$ and the kurtosis $\mu_4$ are defined as
\begin{eqnarray}
 \mu_1 & = & \frac{1}{N^2}\sum_{ij}\kappa_{i,j}\\
 \mu_2 & = & \frac{1}{N^2}\sum_{ij}\left(\kappa_{i,j}-\bar{\kappa}\right)^2\\
 \mu_3 & = & \frac{\mu^{-3/2}_2}{N^2}\sum_{ij}\left(\kappa_{i,j}-\bar{\kappa}\right)^3\\
 \mu_4 & = & \frac{\mu^{-2}_2}{N^2}\sum_{ij}\left(\kappa_{i,j}-\bar{\kappa}\right)^4-3\;,
\end{eqnarray}
where $\bar{\kappa}\equiv \mu_1$ is the mean value of the effective convergence. To evaluate the different moments of 
the convergence maps, we subtract the mean value $\bar{\kappa}$ from the maps, and divide by the total number of 
pixels $N^2$ to get the appropriate normalization.

\begin{figure*}
 \centering
 \includegraphics[width=0.3\textwidth,angle=-90]{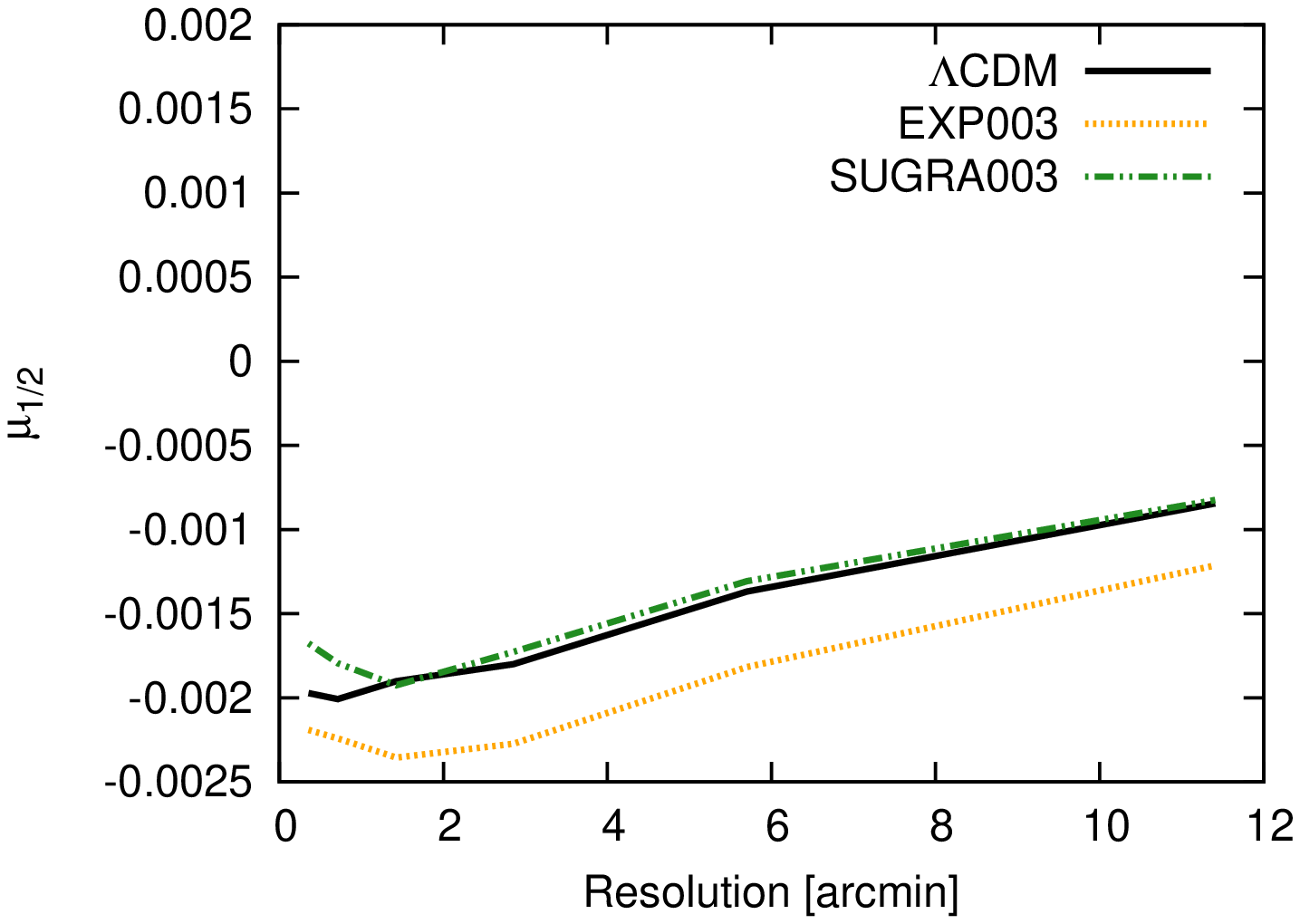}
 \includegraphics[width=0.3\textwidth,angle=-90]{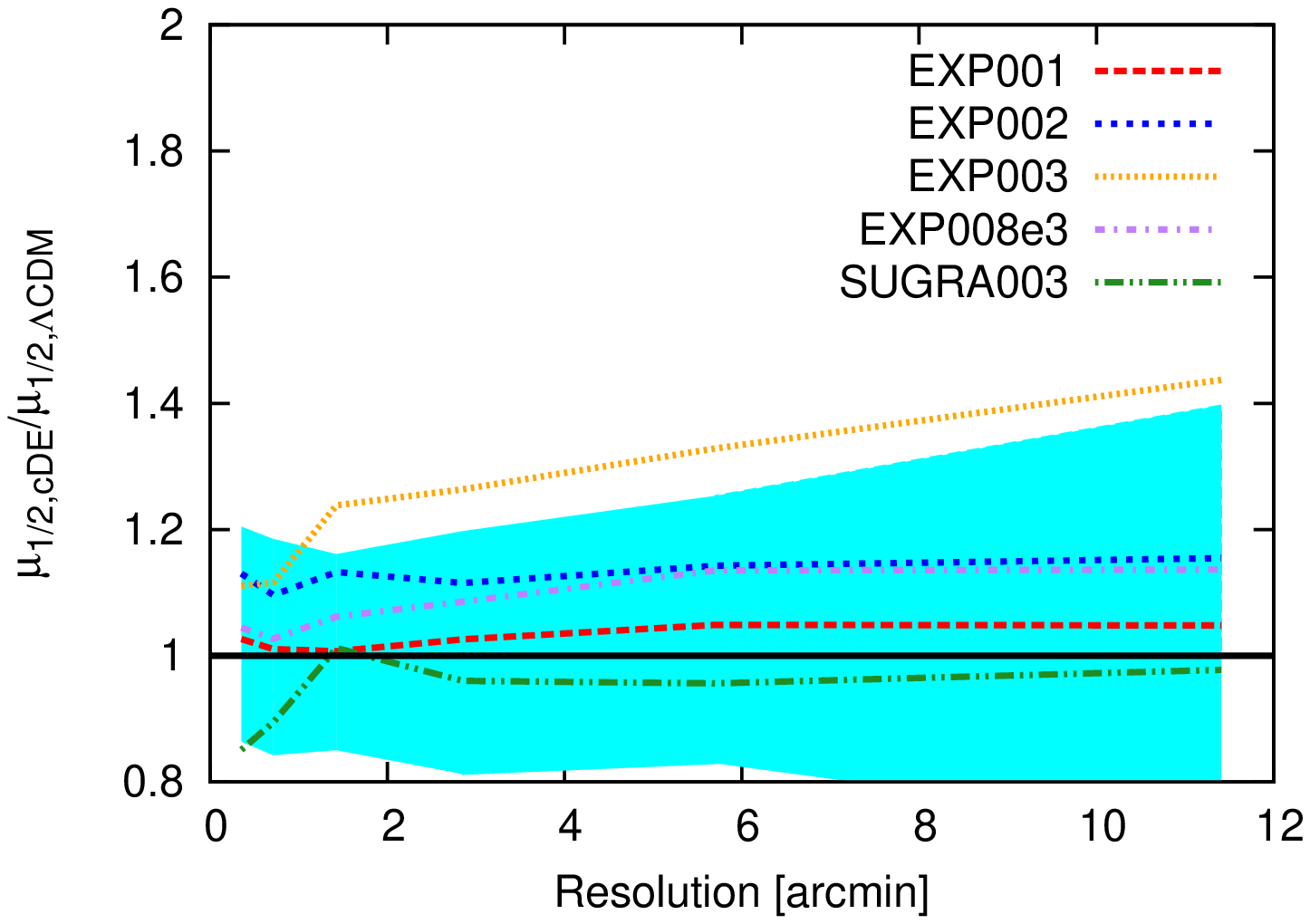}
 \includegraphics[width=0.3\textwidth,angle=-90]{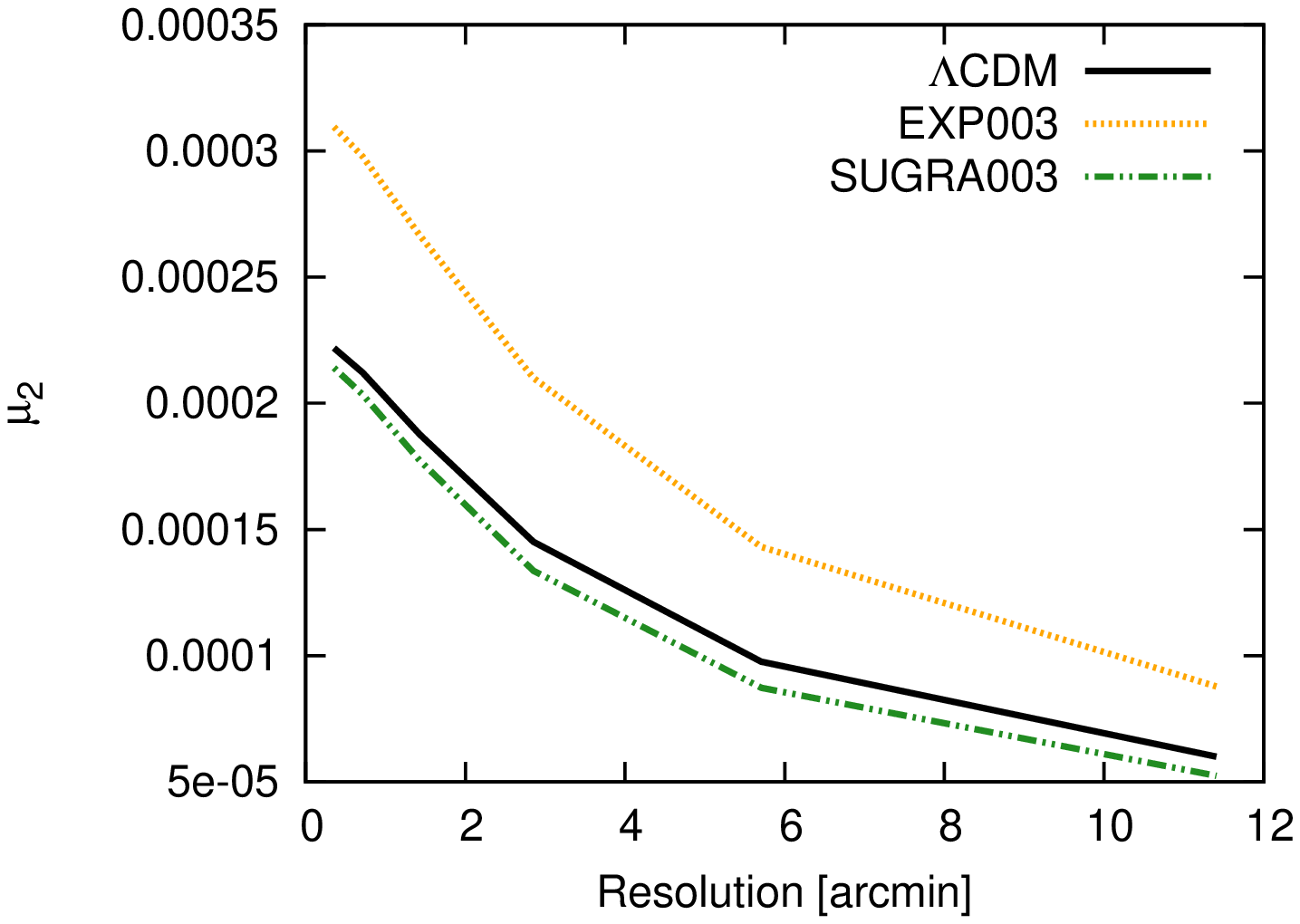}
 \includegraphics[width=0.3\textwidth,angle=-90]{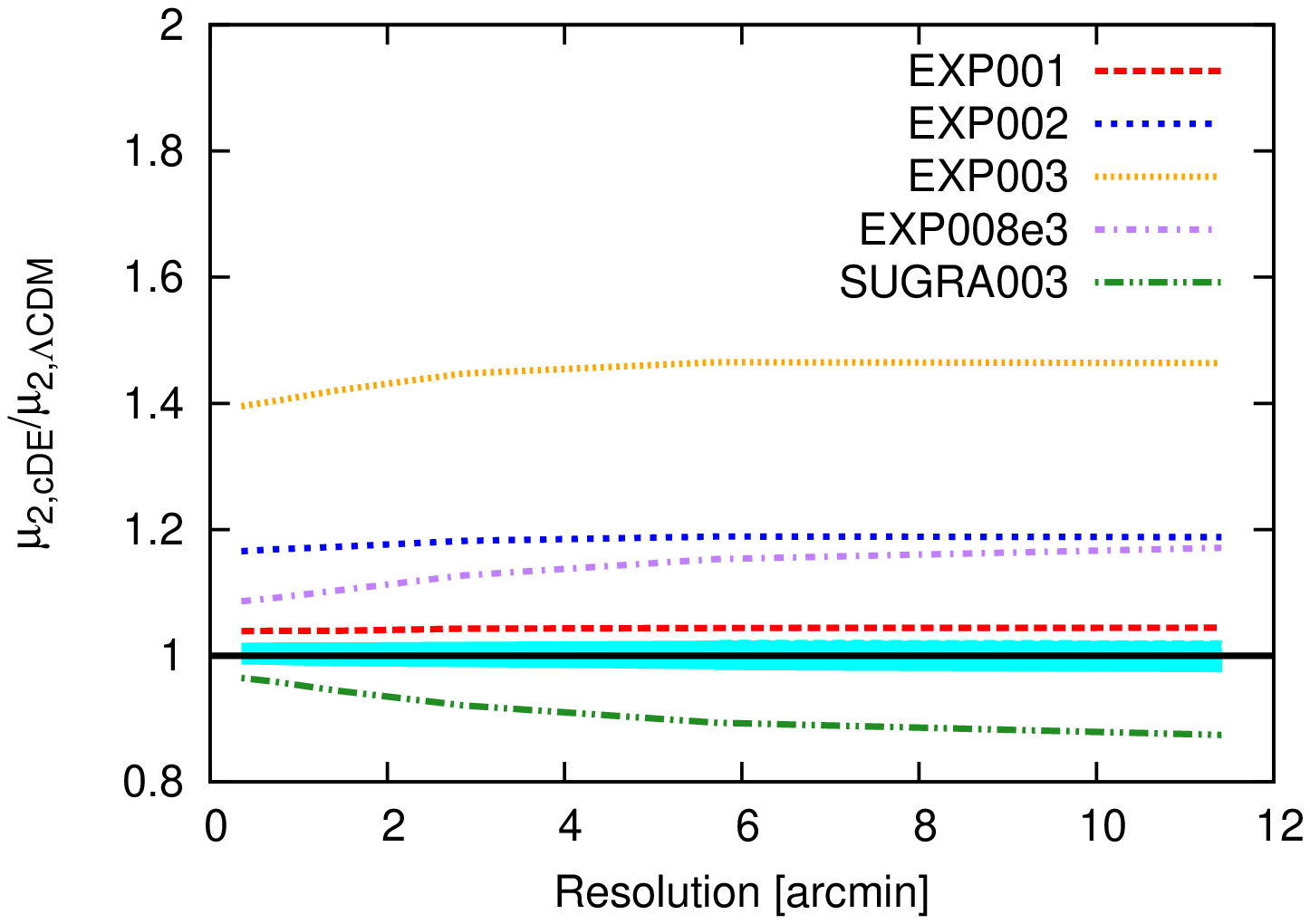}
 \includegraphics[width=0.3\textwidth,angle=-90]{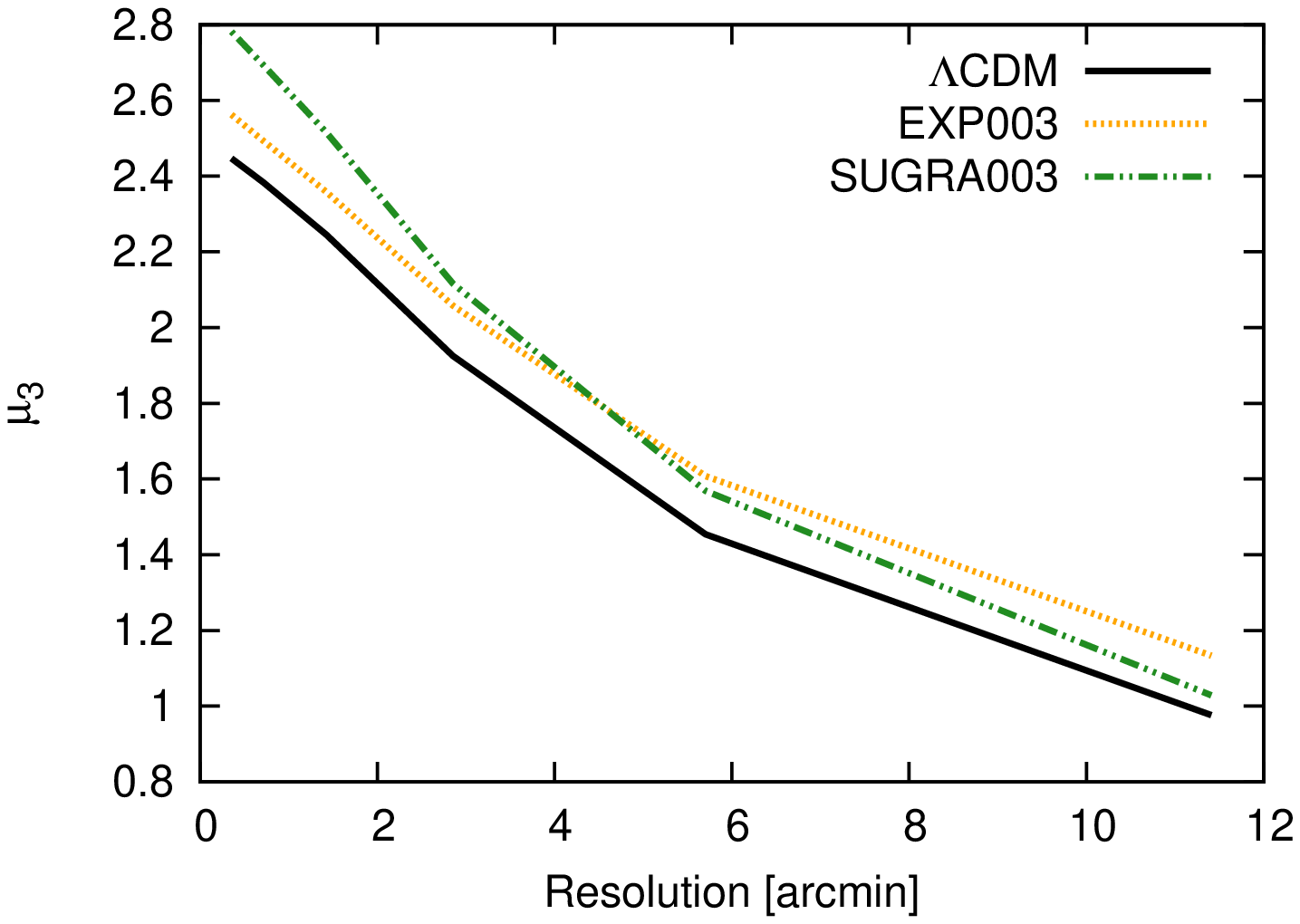}
 \includegraphics[width=0.3\textwidth,angle=-90]{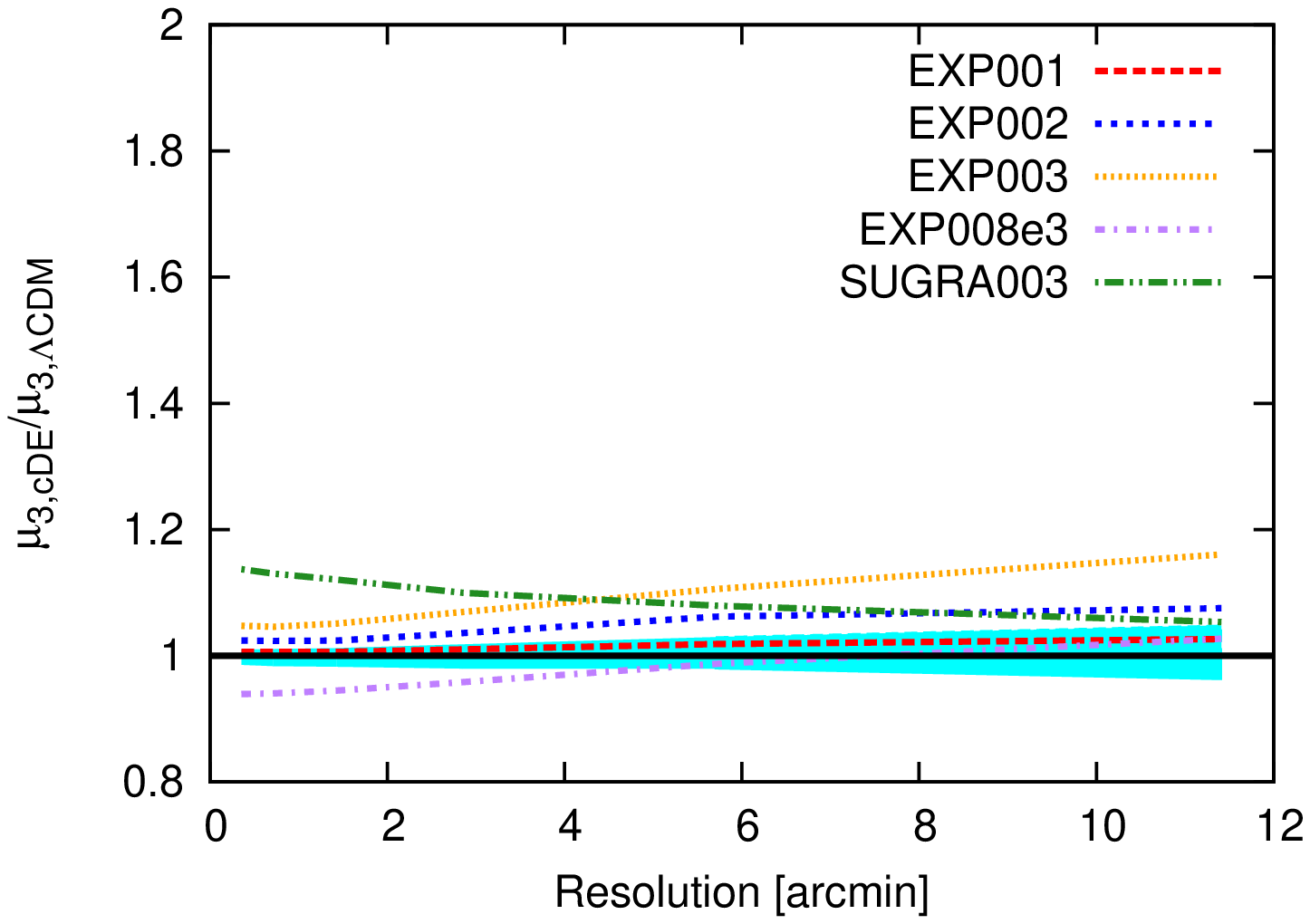}
 \includegraphics[width=0.3\textwidth,angle=-90]{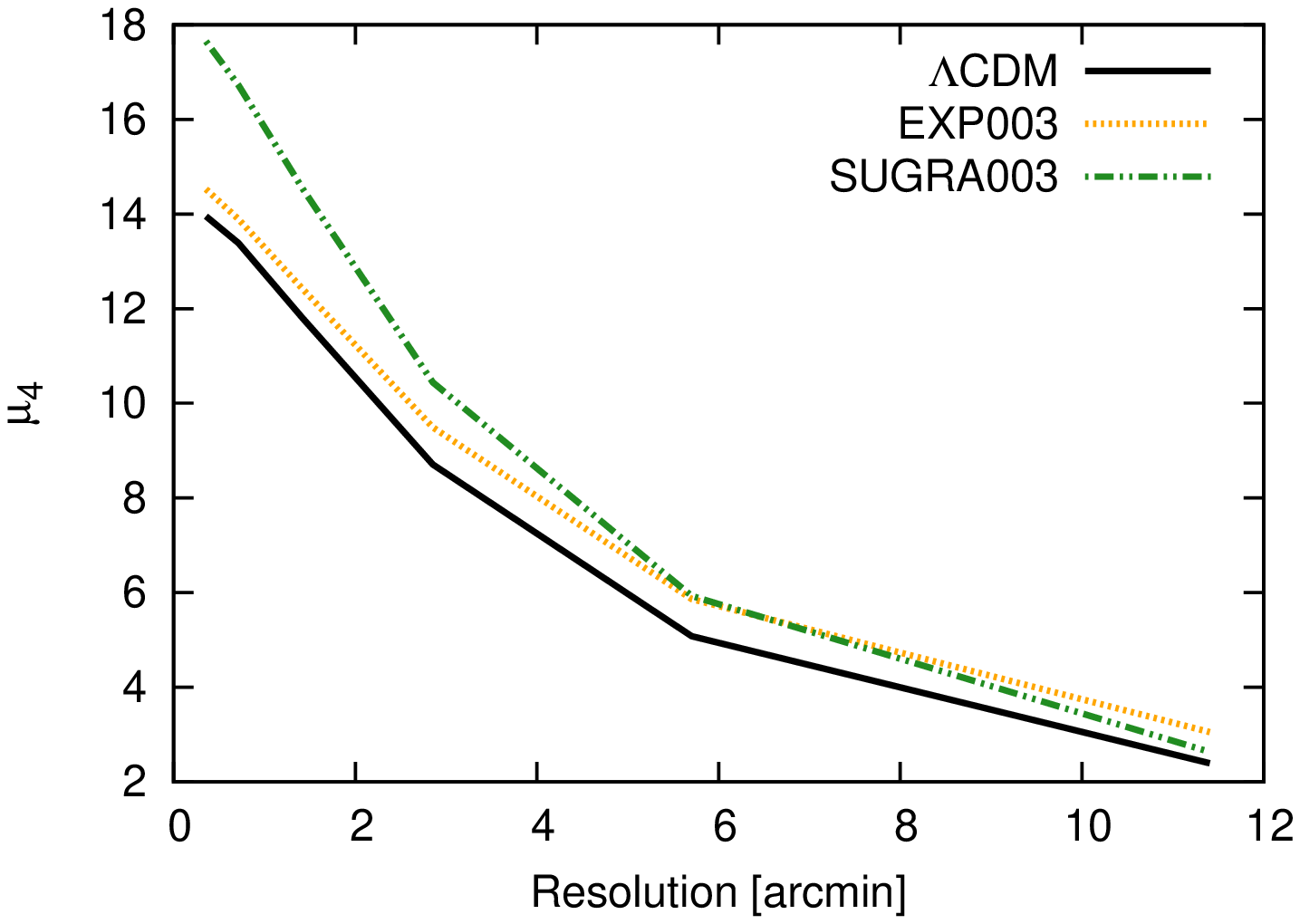}
 \includegraphics[width=0.3\textwidth,angle=-90]{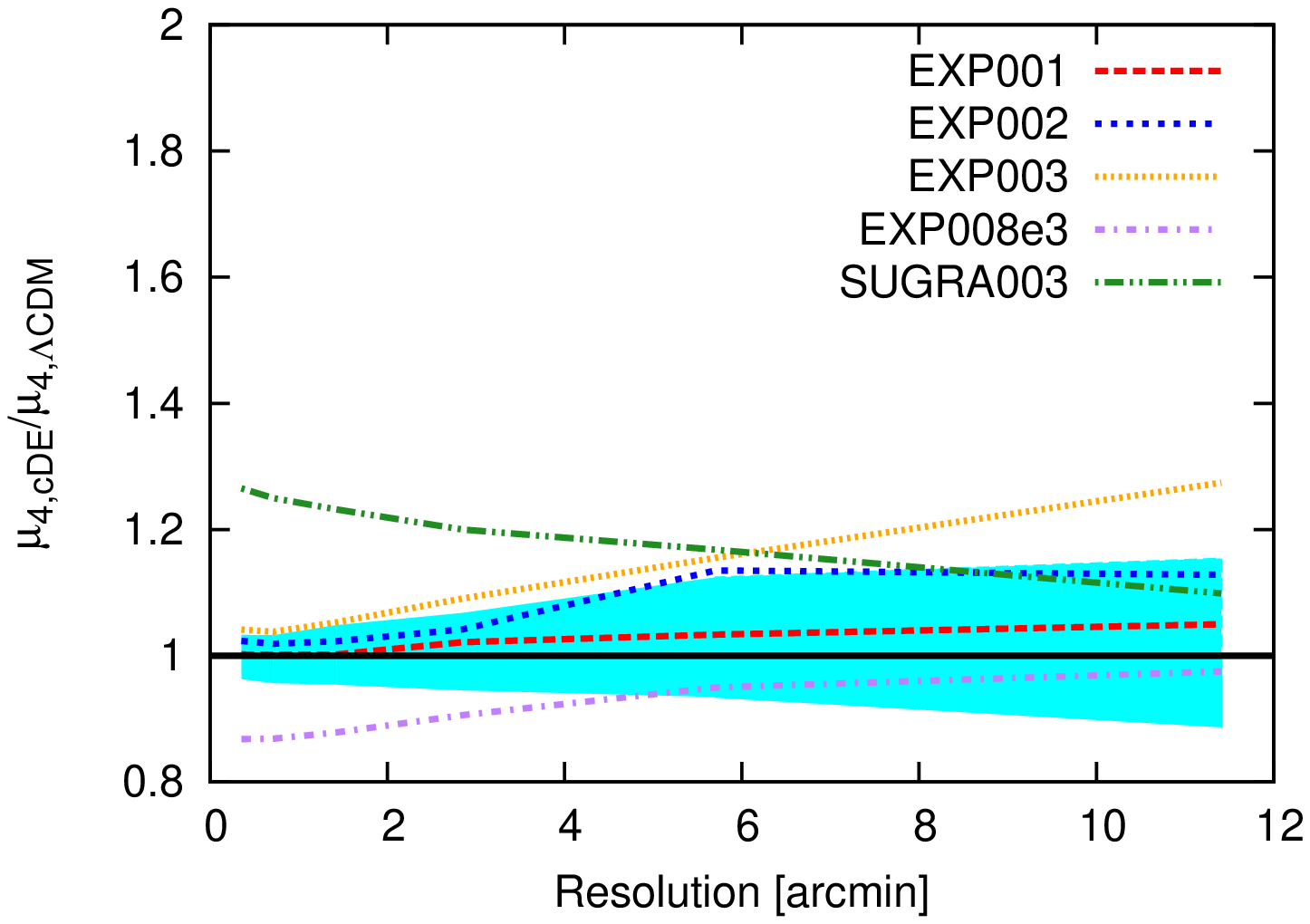}
 \caption{From top to bottom: median of the distribution for the median ($\mu_{1/2}$), variance ($\mu_2$), skewness 
($\mu_3$) and kurtosis ($\mu_4$) of the effective convergence field as a function of the pixel resolution scale. 
Left panels show the results for the reference $\Lambda$CDM model and for the two most extreme coupled dark energy 
models (SUGRA003 and EXP003). Right panels show the ratio between the coupled dark energy models and the $\Lambda$CDM 
model. Colour lines and styles are as in Fig.~\ref{fig:PSkappa}. The curves and the shaded region (shown only for 
$\Lambda$CDM model) represent the median and the quartiles obtained from 100 different realizations, respectively.}
 \label{fig:Resolution}
\end{figure*}

Since the distribution of the convergence is non-Gaussian and its mean is effectively zero, it is useful 
to consider the median, $\mu_{1/2}$, i.e. the value at which the integrated probability is the same above and below. 
In Fig.~\ref{fig:Resolution}, we show the median, the variance, the skewness and the kurtosis as a function of the 
map resolution. To do so, we binned our high resolution convergence maps to progressively decrease the number of 
pixels in the maps, and, as consequence, to make the map resolution progressively worse. Working with sources at 
$z_{\rm s}=1$, we created new sets of maps, with 2048$^2$, 1024$^2$, 512$^2$, 256$^2$ and 128$^2$ pixels. The 
corresponding resolutions are (for a $\Lambda$CDM model), from 4096$^2$ to 128$^2$ pixels, 0.356 arcmin, 0.71 arcmin, 
1.4 arcmin, 2.9 arcmin, 5.7 arcmin and 11.4 arcmin, respectively.

For each quantity we show its median value and the shaded region represents the range between the first and the third 
quartiles of the set of points for the $\Lambda$CDM simulation. Thus it envelopes the central 50\% of the 
distribution, as opposed to the 1-$\sigma$ regions shown previously.

\begin{figure*}
 \centering
 \includegraphics[width=0.3\textwidth,angle=-90]{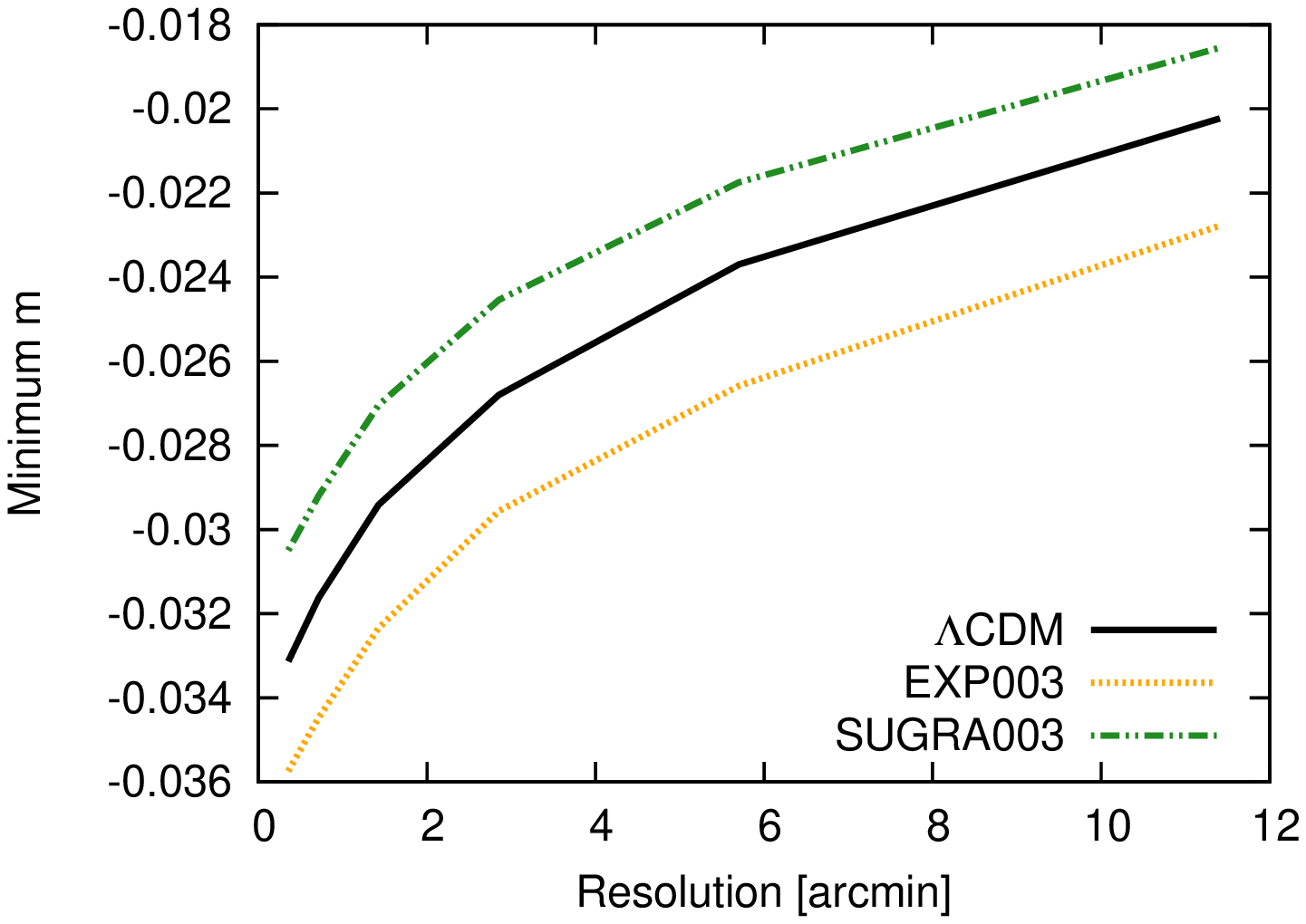}
 \includegraphics[width=0.3\textwidth,angle=-90]{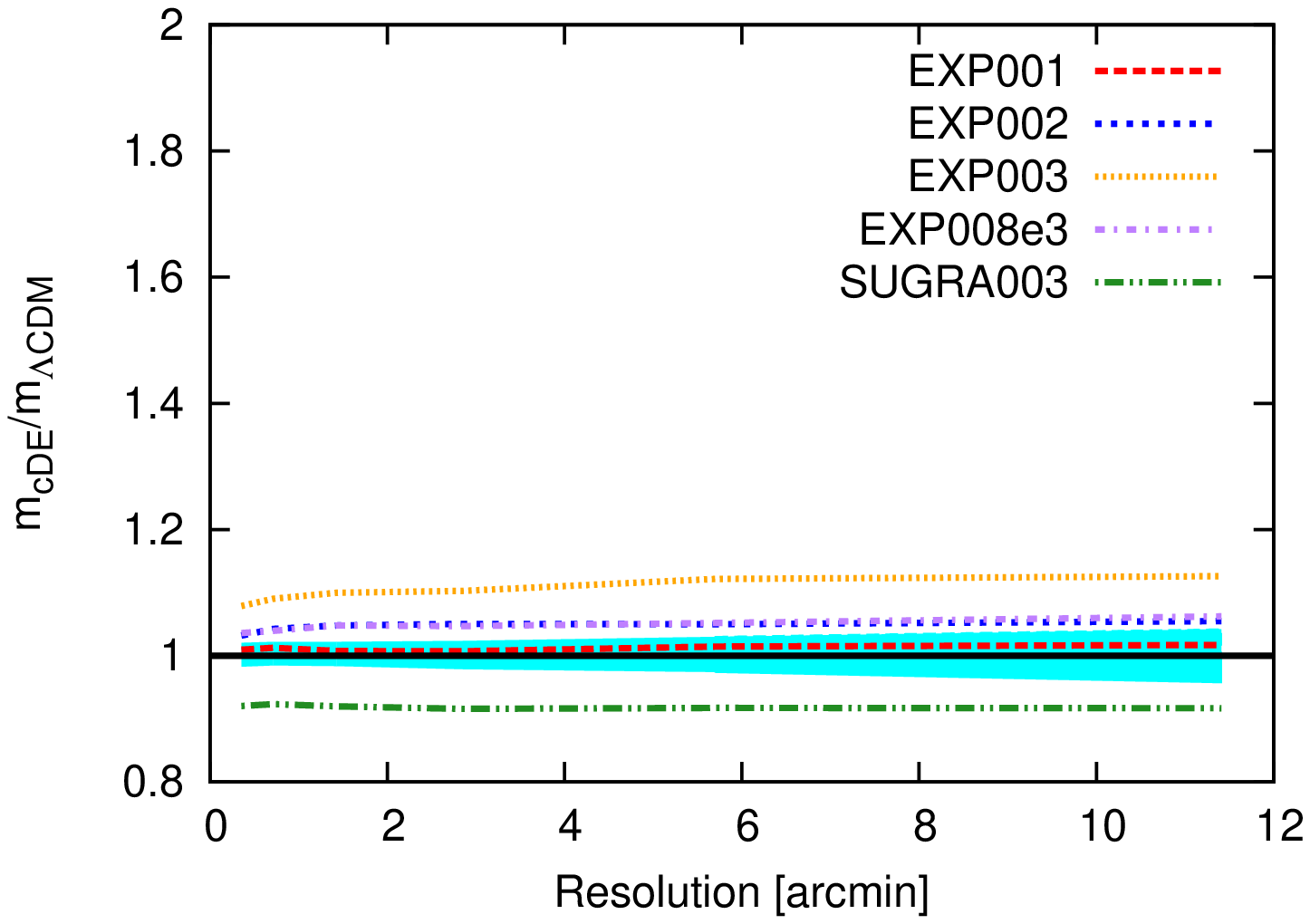}
 \includegraphics[width=0.3\textwidth,angle=-90]{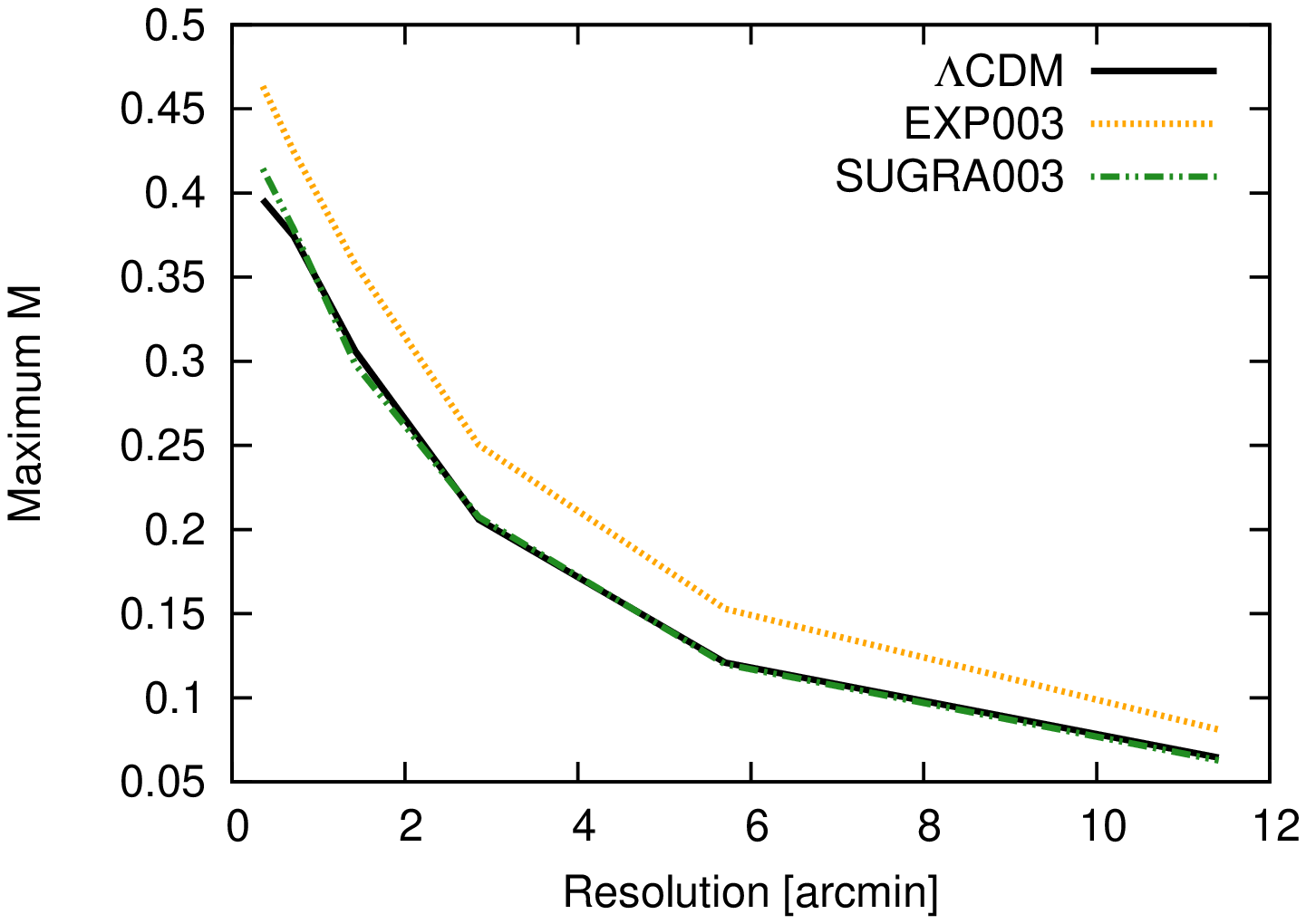}
 \includegraphics[width=0.3\textwidth,angle=-90]{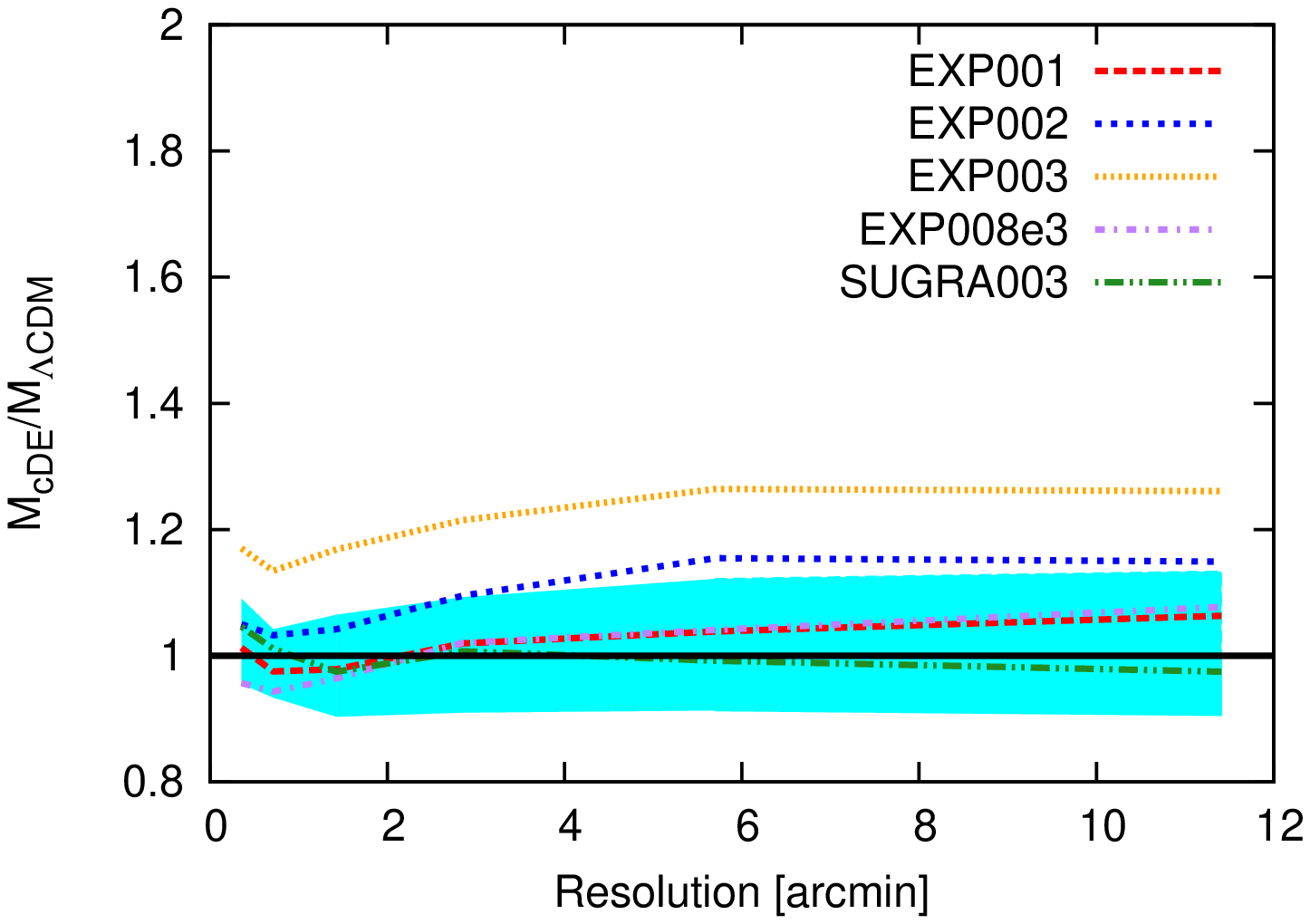}
 \caption{Upper (lower) panels: median of the distribution of minima $m$ (maxima $M$) as a function of the map 
resolution scale. Left panels show the results for the reference $\Lambda$CDM model and for the two most extreme 
coupled dark energy models (SUGRA003 and EXP003). Right panels show the ratio between the coupled dark energy models 
and the $\Lambda$CDM model. Colour lines and styles are as in Fig.~\ref{fig:PSkappa}. The curves and the shaded 
region (shown only for $\Lambda$CDM model) represent the median and the quartiles obtained from 100 different 
realizations, respectively.}
 \label{fig:Resolution1}
\end{figure*}

As expected, the variance shows a very similar behaviour to the convergence power spectrum and shear in aperture, 
also from a quantitative point of view. All the EXP models show higher values for the variance, while the expected 
value for the SUGRA003 model is $\approx 12\%$ lower than the $\Lambda$CDM model. The EXP001 is just outside of the 
quartile area, making therefore difficult to distinguish it at a 1-$\sigma$ level. Other models instead show 
progressively higher differences. As expected, increasing the order of the moments makes the quartiles increase, to 
the point that all the models will be indistinguishable from the reference one. In particular, while for the variance 
only the EXP001 model is comparable with the quartiles, for the skewness only the EXP003 model is more than 
1-$\sigma$ away than the $\Lambda$CDM model, and for the kurtosis all the models are basically within the error bars. 
At high resolution the skewness has the potential of distinguishing between the different models, but its predictive 
power decreases at lower resolutions.

In the case of the median, the area enclosed by the quartiles is rather large, making this statistical quantity 
largely insensitive to the background model, with the exception once again of the EXP003 model. While the median 
changes by a factor of two over the resolution scale analysed in this work, we see that the ratio is approximately 
constant.

We conclude therefore that only the variance can be used as a discriminant between the different models, since for 
higher order statistics, error bars overcome the inner differences between the models. A further comment has to be 
made regarding the error bars and the possibility of using higher order moments of the effective convergence. Error 
bars and quartiles represent effectively the variance between the maps, each of them covering an area of roughly 600 
square degrees. Therefore our conclusions and the possibility of using the skewness and the kurtosis for lensing 
studies are limited to surveys of this size. Larger surveys will have reduced errors bars and higher order moments 
could be used as useful cosmological probes.

Finally, we investigate whether the most extreme behaviour, namely the maximum or minimum value in the entire map, 
might be a good discriminant of the models. In Fig.~\ref{fig:Resolution1} we present the median of the maxima (M) 
and of the minima (m) for the ensemble of effective convergence maps. As expected, these are moderate when the 
resolution decreases and more pixels are averaged together. These highlight the asymmetry of the distributions, as 
the minima are significantly smaller in magnitude compared to the maxima. However, the differences between models are 
quite limited for the minima, at most 10\%-15\% with larger differences for the SUGRA003 and EXP003 models. 
In particular the SUGRA003 (EXP003) model shows less (more) pronounced minima with respect to the $\Lambda$CDM model 
and this can be explained with the different matter density evolution (normalization of the matter power spectrum). 
Maxima instead show a clear trend with normalisation of the matter power spectrum: the higher $\sigma_8$, the higher 
are the differences (up to $\approx 20\%$). It is also worth noting that the distribution of the maxima is very 
sensitive to the map resolution: while minima change only by a factor 1.7, maxima change by a factor of 10. 
As these are rare events, the intrinsic scatter is large, making these a poor discriminator of models.

\section{Conclusions}\label{sect:conclusions}
In this work we have studied weak lensing statistics of coupled dark energy models \citep{Baldi2012}, characterised 
by an interaction between the dark matter and the dark energy. Our aim was to extend previous work on the subject 
\citep[see][]{Beynon2012,Carbone2013}, going beyond the Born approximation with full raytracing simulations to provide 
a realistic simulated suite for lensing quantities, in particular effective convergence, shear, flexions and 
magnification. The advantage of the numerical approach is that full non-linearity is automatically 
achieved and no approximation is necessary for a full analysis (which is usually required with analytical techniques).

A coupling between dark matter and dark energy has important effects on structure formation due to the different 
non-linear evolution of dark matter particles, and the appearance of a fifth force term that, because of its 
frictional nature, tends to suppress non-linear power.

We saw that all the statistical quantities analysed in this work faithfully reproduce features observed in the 
study of the three-dimensional matter distribution. In particular, we observe that:
\begin{itemize}
 \item The effective convergence (shear) power spectrum faithfully reproduces results from \cite{Baldi2012} regarding 
the three-dimensional matter power spectrum. Differences on large scales can be explained by the different 
normalization of the matter power spectrum, but a comparison with a $\Lambda$CDM model having the 
same normalization of the matter power spectrum reveals the importance of the different non-linear evolution, showing 
a suppression of power at small scales. Differences for the coupled dark energy model characterised by a SUGRA 
potential can be explained by the different evolution of the matter density parameter.
 \item PDFs are sensitive to the different background models and could be used to discriminate between the different 
coupled dark energy models. We showed that differences between the models can be mainly explained by differences in 
the normalization of the matter power spectrum, but the high convergence tail can signal differences in the non-linear 
evolution arising from friction terms.
 \item When evaluating the moments of the effective convergence, we find that only the variance can be used as a 
statistical tool to infer the background cosmological model. Higher order statistics like the skewness and the 
kurtosis are more prone to sample variability between the different realizations, making them less 
sensitive for discriminating between the different models.
\end{itemize}

Our simulations have assumed that all the sources are at a fixed redshift, in order to make raytracing 
simulations numerically less expensive. 
The errors we infer are limited by the finite size of the simulations, and would correspond to a moderate sized 
survey of order 600 square degrees, significantly smaller than ongoing or future surveys such as DES or Euclid. Our 
primary aim has been to study whether in principle other weak lensing statistics can provide a useful probe to models 
of coupled dark energy; at the same time, we have developed techniques that will be required to take into 
account non-linear effects in weak lensing.

To conclude, differences on large scales between the coupled dark energy models and the 
$\Lambda$CDM model can largely be explained by the modified growth rate and dark matter fraction, leading to different 
normalisation of the matter 
power spectrum.  On small scales where non-linear effects kick in, a suppression of power is caused by friction 
terms which lead to observable signatures in the power spectrum and the probability distribution function.

\section*{Acknowledgements}
Raytracing simulations were run on the Intel SCIAMA High Performance Compute (HPC) cluster which is supported by the 
ICG, SEPNet and the University of Portsmouth. F.~P., D.~B. and R.~C. are supported by STFC grant ST/H002774/1.
M.~B. is supported by the  Marie Curie Intra European Fellowship ``SIDUN"  within the 7th Framework  Programme of 
the European Commission.
M.~B. and L.~M. acknowledge financial contributions from contract ASI/INAF I/023/12/0, from PRIN MIUR 2010-2011 ”The 
dark Universe and the cosmic evolution of baryons: from current surveys to Euclid” and from PRIN INAF 2012 
”The Universe in the box: multi-scale simulations of cosmic structure”. F.~P. thanks Andrea Macci\`o for discussions 
at an early stage of the project.\\
The authors also thank the anonymous referee whose comments helped us to improve the presentation of our results.

\bibliographystyle{mn2e}
\bibliography{WL_CoDECS.bbl}

\label{lastpage}

\end{document}